\documentclass{emulateapj}
\usepackage{natbib,amsfonts,amsmath,amssymb,rotating,lscape}

\newcommand{\about}     {\hbox{$\sim$}}

\newcommand{\gM}        {\emph{g}}

\newcommand{\iM}        {\emph{i}}

\newcommand{\si}[1]{\ensuremath{_{\textrm{\scriptsize{#1}}}}}

\begin{document}

\title{Soft X-ray and Ultraviolet Emission Relations in Optically Selected AGN Samples}

\author{
Iskra V. Strateva\altaffilmark{1},
W. N. Brandt\altaffilmark{1},
Donald P. Schneider\altaffilmark{1},
Daniel G. Vanden Berk\altaffilmark{1},
Cristian Vignali\altaffilmark{2,3}
}

\altaffiltext{1}{Department of Astronomy and Astrophysics, 525 Davey Lab, 
Pennsylvania State University, University Park, PA 16802} 
\altaffiltext{2}{Dipartimento di Astronomia, Universit\`a degli Studi di Bologna,
Via Ranzani 1, 40127 Bologna, Italy}
\altaffiltext{3}{INAF -- Osservatorio Astronomico di Bologna, Via Ranzani 1,
40127 Bologna, Italy}

\addtocounter{footnote}{3} 

\begin{abstract}
Using a sample of 228 optically selected Active Galactic Nuclei (AGNs)
in the \hbox{0.01--6.3} redshift range with a high fraction of X-ray
detections (81--86\%), we study the relation between rest-frame UV and
soft X-ray emission and its evolution with cosmic time. The majority
of the AGNs in our sample (155 objects) have been selected from the
Sloan Digital Sky Survey (SDSS) in an unbiased way, rendering the
sample results representative of all SDSS AGNs. The addition of two
heterogeneous samples of 36 high-redshift and 37 low-redshift AGNs
further supports and extends our conclusions.  We confirm that the
X-ray emission from AGNs is correlated with their UV emission, and
that the ratio of the monochromatic luminosity emitted at 2\,keV
compared to 2500\,\AA\ decreases with increasing luminosity
(\hbox{$\alpha\si{ox}=-0.136 l\si{uv}+2.616$}, where $l\si{uv}$ is in
$\log$ units), but does not change with cosmic time.  These results
apply to intrinsic AGN emission, as we correct or control for the
effects of the host galaxy, UV/X-ray absorption, and any X-ray
emission associated with radio emission in AGNs. We investigate a
variety of systematic errors and can thereby state with confidence
that (1) the $\alpha\si{ox}$--$l\si{uv}$ anti-correlation is real and
not a result of accumulated systematic errors and (2) any
$\alpha\si{ox}$ dependence on redshift is negligible in comparison. We
provide the best quantification of the $\alpha\si{ox}$--$l\si{uv}$
relation to date for normal radio-quiet AGNs; this should be of
utility for researchers pursuing a variety of studies.
\end{abstract}    


\section{Introduction}
\label{intro}

Surveys for Active Galactic Nuclei\footnote{In this paper ``AGNs''
refers to all types of active galaxies covering the full range of
observed luminosities.}  (AGNs) were until recently most commonly
conducted in the observed optical band (corresponding to the
rest-frame UV for high-redshift AGNs); consequently, our understanding
of the AGN population is biased toward properties inferred from AGN
samples bright in the optical. Radio, infrared, and X-ray surveys have
revealed more reddened and obscured AGNs, attesting to the presence of
an optical bias. AGN surveys in non-optical bands still require
optical or UV spectroscopy to confirm the presence of an active
nucleus (except for bright, hard X-ray selected AGNs, or AGNs with
large radio jets) and to determine the redshift. Historically, our
understanding of the evolution of the luminous AGN population with
cosmic time has been based largely on optically selected AGN samples;
use of samples selected in other bands to further this understanding
requires proper interpretation of the relations between emission in
these bands and optical/UV emission for comparison. X-ray surveys are
more penetrating and efficient in separating the host-galaxy
contribution from the nuclear emission for sources with
$L\si{x}\gtrsim10^{42}$\,erg\,s$^{-1}$, as the integrated host-galaxy
X-ray emission is negligible compared to the nuclear emission (which
contributes \hbox{5--30\%} of the AGN bolometric luminosity). In order
to compare X-ray survey results on AGN evolution to those in the
optical/UV, as well as to understand better the details of the nuclear
environment and the accretion process powering AGNs, we need to
establish the relations between optical/UV and X-ray emission in
optically selected samples.

\citet{Tananbaum79} discovered that a large fraction of UV-excess and 
radio-selected AGNs are strong X-ray sources with X-ray luminosities
correlated with those measured in the rest-frame UV. This result was
confirmed by \citet{Zamorani81}, who also found that the X-ray
emission of AGNs depends on their radio power (with radio-loud AGNs
being on average $\approx3$ times brighter in X-rays) and that the
optical/UV-to-X-ray monochromatic flux ratios of AGNs depend on rest-frame
UV luminosity and/or redshift. The relation between AGN emission in
the rest-frame UV and X-ray bands is commonly cast into a ratio of
monochromatic fluxes called ``optical/UV-to-X-ray index'', $\alpha\si{ox}$,
defined as the slope of a hypothetical power law extending between
2500\,\AA\ and 2\,keV in the AGN rest frame\footnote{The subscript of
$\alpha\si{ox}$ comes from the name ``optical-to-X-ray
index''. ``Optical'' is somewhat of a misnomer since it refers to the
ultraviolet (2500\,\AA\ rest-frame) monochromatic flux which falls in the
observed optical band for most bright AGNs studied originally. We use
``optical/UV-to-X-ray index'' instead but retain the designation
$\alpha\si{ox}$ for historical reasons.}:
$\alpha\si{ox}=-0.3838\log[F_{\nu}(\textrm{2\,keV})/F_{\nu}(\textrm{2500\,\AA})]$.
Studies of optical/UV and radio samples of AGNs observed with the
\emph{Einstein Observatory} \citep[e.g.,][]{AT82,KC85,AT86,AM87,XrayRadio,Wilkes94} 
and \emph{ROSAT} \citep[e.g.,][]{Green95} confirmed that over 90\% of
optically selected AGNs are luminous X-ray emitters, that the X-ray
emission from AGNs (from Seyfert 1s to luminous QSOs) is correlated
with the optical/UV emission as well as the radio emission, and that
the primary $\alpha\si{ox}$ dependence is most likely on optical/UV
luminosity rather than redshift \citep[but
see][]{Yuan98,Bechtold03}. The most comprehensive recent study of
X-ray emission from a radio-quiet (RQ) sample of optically selected
AGNs is that of \citet[][hereafter VBS03]{VBS03}, who found a stronger
$\alpha\si{ox}$ dependence on rest-frame UV luminosity than redshift.

A robust empirical study of the relations between optical/UV and X-ray
emission from AGNs provides a valuable basis for theoretical studies
of AGN energy-generation mechanisms. As we discuss in
$\S$\ref{conclude}, there are no concrete theoretical studies to date
predicting the observed range of $\alpha\si{ox}$ or its dependence on
rest-frame UV luminosity and/or redshift. A well-calibrated rest-frame
UV-to-X-ray relation can also be used to derive reliable estimates of
the X-ray emission from optically selected, RQ, unabsorbed AGNs and
can lead to improved bolometric luminosity estimates. Furthermore,
refined knowledge of the ``normal'' range of rest-frame UV-to-X-ray
luminosity ratios in AGNs is necessary to define more accurately
special AGN subclasses (e.g., X-ray weak AGNs) and (under certain
assumptions) estimate the X-ray emission associated with jets in RL
AGNs.

Establishing the relations between the intrinsic rest-frame UV and
X-ray emission in optically selected samples (excluding the effects of
absorption and jet-associated X-ray emission) can be done efficiently
and accurately only with samples with a high fraction of X-ray
detections, optical/UV spectroscopy, and radio classifications. In
addition, appropriate statistical-analysis methods developed to detect
partial correlations in censored data sets must be used.  The advent
of large-area, highly complete optical surveys like the 2 degree Field
Survey \citep[2dF,][]{2dF} and the Sloan Digital Sky Survey
\citep[SDSS; ][]{York00}, coupled with the increased sky coverage of
medium-depth X-ray imaging (pointed observation with the
\emph{ROentgen SATellite} -- \emph{ROSAT}, \emph{X-ray Multi-Mirror
Mission-Newton} -- \emph{XMM-Newton}, and \emph{Chandra X-ray
Observatory} -- \emph{Chandra}), make the task of creating suitable
samples feasible.  We have constructed a sample of 155 SDSS AGNs in
medium-deep \emph{ROSAT} fields, supplemented with a low-redshift
Seyfert~1 sample and a high-redshift luminous AGN sample (for a total
of 228 AGNs), to investigate the relation between rest-frame UV and
soft X-ray emission in RQ AGNs.  Several important conditions must be
met to ensure the appropriateness of the sample and statistical
methods:

1. Large ranges of luminosity and redshift must be sampled to reveal
weak correlations of $\alpha\si{ox}$ with luminosity and
redshift. Additionally, a significant range in luminosity at each
redshift is necessary to control for the strong redshift dependence of
luminosity in flux-limited samples \citep[e.g.,][]{AT86}; this range
should be larger than the observed measurement and variability
dispersions. Our current sample of 228 AGNs covers the largest
redshift and luminosity ranges to date,
\hbox{$0.01<z<6.3$} and $10^{27.5}$\,erg\,s$^{-1}$\,Hz$^{-1}$
$<L$\si{$\nu$}(2500\,\AA)$<10^{33}$\,erg\,s$^{-1}$\,Hz$^{-1}$, without
sacrificing a high X-ray detection fraction or seriously affecting the
sample homogeneity. The main SDSS sample provides adequate luminosity
coverage in the \hbox{$0.2<z<3.0$} redshift range; the addition of the
Seyfert~1 and \hbox{high-$z$} AGN samples (see $\S$\ref{sample})
increases the range of luminosities probed at low and high redshifts,
respectively.

2. It is necessary to determine the radio loudness of each AGN and to
exclude the strongly radio-loud (RL) AGNs. RL AGNs have more complex
mechanisms of energy generation, such as jet emission, which can
obscure the X-ray emission directly associated with accretion
(particularly if an AGN is observed at a small viewing angle). The
Faint Images of the Radio Sky at Twenty-Centimeters survey
\citep[FIRST;][]{first} was designed to cover most of the SDSS
footprint on the sky, providing sensitive 20\,cm detections
(\about1\,mJy -- 5$\sigma$) and limits that allow us to exclude
strongly RL AGNs.  Some previous studies lacked adequate radio
coverage and/or did not separate these two AGN populations.

3. Because we wish to quantify any evolution of the main intrinsic
energy generation mechanism in AGNs, it is necessary to exclude AGNs
strongly affected by absorption. Strong X-ray absorption in AGNs is
often associated with the presence of broad ultraviolet absorption
lines \citep[e.g.,][]{BLW,Gallagher}. The large observed wavelength
range and high signal-to-noise (S/N) of the SDSS spectroscopy is
sufficient to find Broad Absorption Line (BAL) AGNs in 40--70\% of the
sample (see below), allowing us to limit the confusing effects of
X-ray absorption.

4. Special statistical tools are needed to evaluate correlations when
censored data points are present. We use the rank correlation
coefficients method described by \citet{AS96} to determine the
significance of correlations in the presence of censored data points,
while taking into account third-variable dependencies.  Using Monte
Carlo simulations, we confirm the robustness of the correlation
significance estimates. We derive linear regression parameters in two
independent ways, using the Estimate and Maximize (EM) and the
Buckley-James regression methods from the Astronomy SURvival Analysis
package \citep[ASURV;][]{LIF92,IFN85,IFN86}.  

5. In addition to the use of appropriate statistical tools, a large
detection fraction is necessary to infer reliable correlations in
censored data samples. \citet{A85} and \citet{AM87} outline the biases
that can affect the sample means and correlation parameters as a
result of systematic pattern censoring. Our current sample has 86\%
X-ray detections (compared to
\hbox{\about 10-- 50\%} for previous studies).  One of the assumptions
of the statistical methods described in (4), which could be violated,
is that the AGNs with upper limits and detections have the same
underlying distributions of $\alpha\si{ox}$ and rest-frame UV
luminosity. The effect of this assumption is partially alleviated by
excluding RL and BAL AGNs, but achieving a high detection fraction is
the only definitive way to suppress the effect of the unknown and
likely different distributions of $\alpha\si{ox}$ and rest-frame UV
luminosity for AGNs with X-ray detections and limits.

6. The results from statistical analyses must take into account the
findings of \citet{Chanan83}, \citet{LaFranca}, and \citet{Yuan98}
that apparent correlations can be caused by a large dispersion of the
measured monochromatic luminosity in the optical/UV relative to the
X-ray band. In this work we use Monte Carlo simulations of our sample
\citep[as described in][]{Yuan99} to confirm the robustness of the
present correlations.

7. Unlike previous studies, we measure directly the rest-frame UV
monochromatic flux at 2500\,\AA\ in three-quarters of the AGNs
comprising the SDSS sample, which guarantees measurement errors of
$\lesssim$10\%. This is made possible by the improved
spectrophotometry of SDSS Data Release Two
\citep[DR2;][]{sdssDR2}.

8. Special care is needed to account for the effects of host-galaxy
contamination of the rest-frame UV monochromatic flux measurements for
low-luminosity AGNs. The high-quality and large wavelength range of
the SDSS spectra are well suited for this.

9. If several X-ray instruments or reductions are used to measure
X-ray monochromatic fluxes, it is necessary to assess mission-to-mission
cross-calibration uncertainties and the effects of different reduction
techniques. The majority of the objects in our sample come from one
instrument \citep[the \emph{ROSAT} Position Sensitive Proportional
Counter -- \emph{ROSAT} PSPC;][]{RosatRef} and were processed
uniformly (see $\S$\ref{xray}), while cross-mission comparisons
between \emph{ROSAT} and \emph{XMM-Newton} or \emph{Chandra} allow
estimation of the effects of inhomogeneity caused by
mission-to-mission cross-calibration issues.

10. Due to the timing of most previous studies coupled with the recent
precise determination of the cosmological parameters, the
``consensus'' cosmology used for luminosity estimates has changed. In
what follows, we use the \emph{Wilkinson Microwave Anisotropy Probe}
cosmology parameters from \citet{Spergel03} to compute the
luminosities of AGNs: $\Omega_{\Lambda}=0.73$, flat cosmology, with
$H_0$=72\,km\,s$^{-1}$\,Mpc$^{-1}$.

The largest optically selected AGN sample with a high fraction of
X-ray detections ($\gtrsim 50\%$) used for establishing the relations
between optical/UV and X-ray emission to date is the VBS03 sample of
SDSS AGNs in regions of pointed \emph{ROSAT} PSPC observations. The
VBS03 sample consists of \about 140 RQ AGNs from the SDSS Early Data
Release \citep[EDR;][]{edr} with a soft X-ray detection fraction of
\about 50\%, supplemented by higher redshift optically selected
AGNs. The second data release of the SDSS provides a large AGN sample
(\about 9 times that of the EDR) with accurate spectrophotometry,
which together with the large medium-deep \emph{ROSAT} sky coverage,
allows us to improve the VBS03 study significantly by increasing the
detection fraction to $>$80\% for a similar size sample, while taking
into account the effects of host-galaxy contributions in the
optical/UV for lower luminosity, nearby AGNs. In this paper we
consider in detail the correlation between rest-frame UV and soft
X-ray emission in AGNs and the dependence of $\alpha\si{ox}$ on
redshift and rest-frame UV luminosity in a combined sample of 228 AGNs
with no known strong UV absorption or strong radio emission.

\section{Sample Selection and X-ray Flux Measurements}
\label{sample}

As described in detail below, we start with \about35,000 AGNs from the
SDSS DR2 catalog, of which we select 174 AGNs with medium-deep
\emph{ROSAT} coverage in the \hbox{0.5--2\,keV} band. From the initial
sample of 174 AGNs we select 155 by excluding all BAL and strong
radio-emission objects. The X-ray detection fraction of this sample of
155 AGNs is 81\%, and we refer to this set as the ``main'' sample.  We
supplement the SDSS data, which cover the \hbox{$0.2<z<3.5$} redshift
range, with additional high- and low-redshift samples, thereby also
increasing the luminosity range covered at the lowest redshifts. We
note that all of the main results of this study can be obtained from
the main sample alone and are reported separately. The ``high-$z$''
sample consists of 36 AGNs with 31 X-ray detections from
\emph{Chandra} and \emph{XMM-Newton} covering the redshift range
\hbox{$4.0<z<6.3$}. The low-redshift Seyfert~1 sample (hereafter ``Sy
\,1'') consists of 37 AGNs detected with \emph{ROSAT} and the
\emph{International Ultraviolet Explorer} (\emph{IUE}) with
\hbox{$z<0.11$}.  We refer to all AGNs from the main, high-$z$, and
Sy\,1 samples as the ``combined'' sample. The combined sample consists
of 228 AGNs with 195 X-ray detections (86\%).

\begin{figure}
\epsscale{1.0}
\plotone{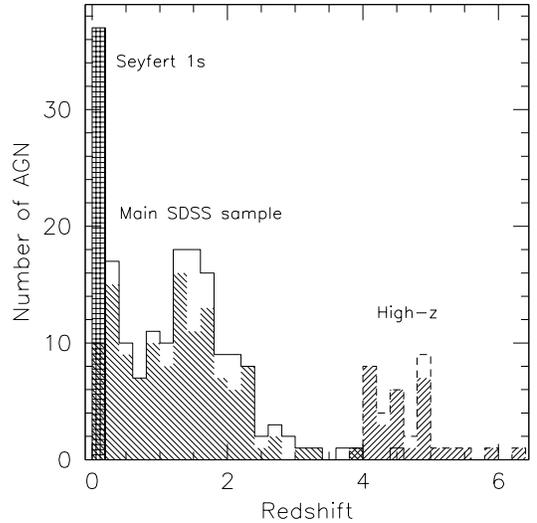}
\caption{The redshift distributions of the main SDSS sample 
(solid-line histogram), the high-$z$ sample (dashed-line histogram),
and the Seyfert 1 sample (block-hatched histogram). The hatched part
of each histogram denotes the AGNs with X-ray detections.
\label{zHist}} 
\end{figure}

The redshift distributions of the main, high-$z$, and Sy\,1 samples
are presented in Figure~\ref{zHist}. High-redshift AGNs are relatively
rare \citep[e.g., see the SDSS DR1 AGN catalog; ][]{QSOcat}, and
consequently there are only eight $z>3$ AGNs in medium-deep
\emph{ROSAT} pointings in our main sample.  The median redshift of the
main SDSS sample is $z\si{median}=1.3$, compared to $z\si{median}=4.5$
for the high-$z$ sample, and $z\si{median}=0.035$ for the Sy\,1
sample.

\begin{figure}
\epsscale{1.0}
\plotone{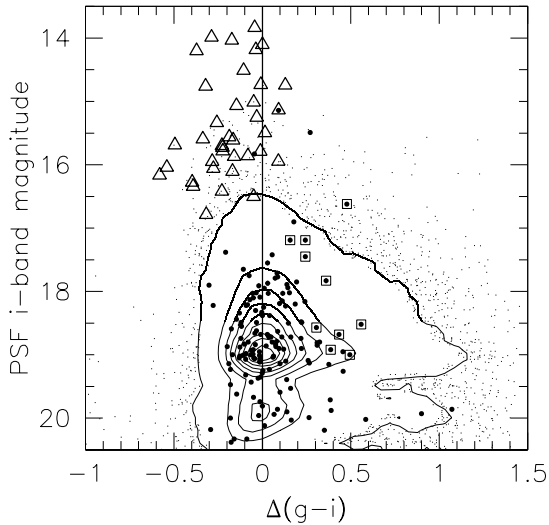}
\caption{Relative \gM$-$\iM\ color, $\Delta(g-i)$, vs.\ apparent \iM-band 
PSF magnitude for the main SDSS sample (solid circles) in comparison
with the full SDSS DR2 sample (linearly spaced contours enclosing 90\%
of the data and small dots representing the outliers) and 37 AGNs from
the BQS sample with SDSS coverage in DR3 (open triangles). The ten
main SDSS sample AGNs, whose relative colors are additionally reddened
by their host galaxies, have their symbols enclosed by large open
squares.  Note that the main SDSS sample is representative of SDSS
AGNs in general and that it covers a different range of colors than
the BQS, as shown for the full SDSS AGN sample by \citet{Jester}. In
the computation of $\Delta(g-i)$, only AGNs with point-source
morphology were used to determine the median \gM$-$\iM\ color as a
function of redshift to prevent artificial reddening due to
host-galaxy contamination. This results in poor sampling and increased
errors at $z<0.08$, affecting 12 of the BQS AGNs. The four $z>3.1$
main-sample AGNs with $\Delta(g-i)$ values affected by the Lyman limit
are excluded.
\label{sdssBQS}} 
\end{figure}

\subsection{SDSS Optical AGN Selection}
\label{optical}
The SDSS \citep{York00} is an imaging and spectroscopic survey with
the ambitious goal of covering a quarter of the celestial sphere,
primarily at the Northern Galactic Cap. AGNs are targeted for
spectroscopy based on a four-dimensional color-selection algorithm
which is highly efficient and able to select AGNs redder than
traditional UV-excess selection surveys
\citep{Richards02,Richards03,Hopkins}.  Assuming that \about 15\% of
the AGN population is reddened, SDSS target selection recovers about
40\% of these reddened AGNs (G. Richards 2004, private
communication). Figure~\ref{sdssBQS} displays the apparent \iM-band
Point Spread Function (PSF) magnitude vs. relative \gM$-$\iM\ color,
$\Delta(g-i)$, constructed by subtracting the median \gM$-$\iM\ color
of DR2 AGNs as a function of redshift from each observed AGN color in
our main sample \citep{Richards03}. This plot was inspired by
\citet{Jester}, who show that the SDSS AGN survey includes objects
with a much wider range of \hbox{$g-i$} colors than the brightest
$B$-band selected AGNs (even at comparable $i$-band magnitudes),
suggesting that popular samples such as the Bright Quasar Survey
sample \citep[hereafter BQS;][]{BQS} might not be representative of
larger and fainter AGN samples with red-band flux cuts like the
SDSS.\footnote{At low redshift, intrinsically faint AGNs will have
redder colors in comparison to bright AGNs due to larger host-galaxy
contributions, even when PSF magnitudes are used to estimate the
relative color. This could affect 10 AGNs from the main sample which
have substantial host-galaxy contributions (as estimated by their
3$''$-aperture spectrum at the end of this section), but it will not
affect significantly the BQS AGNs.}  Figure~\ref{sdssBQS} shows that
our main SDSS sample is representative of SDSS AGNs in general and
contains substantially redder AGNs than 37 BQS AGNs contained in the
SDSS Data Release~3 \citep[DR3;][]{sdssDR3} coverage (four additional
BQS AGNs, whose images are saturated in the SDSS exposures, are
omitted from this plot). This color difference is caused in part by
the shallow $B$-band cut of the BQS survey (sampling of fainter AGNs
reveals both redder and bluer AGNs, as the broadening of the
$\Delta(g-i)$ distribution with fainter \iM\ shows in
Figure~\ref{sdssBQS}), as well as the blue-band selection and blue
\hbox{$U-B$} cut of the BQS \citep{Jester}.

We ensure that all SDSS AGNs considered here were targeted as one of
the QSO target subclasses \citep{edr,Richards02}, excluding objects
targeted solely as FIRST or \emph{ROSAT} sources. The efficiency of
the SDSS target selection (spectroscopically confirmed AGNs as
fraction of targets) is \about 66\%, while the estimated completeness
(fraction of all AGNs above a given optical flux limit in a given area
that are targeted) is \about95\% for point sources with $\iM<19.1$,
which drops to \about 60\% for the high-redshift selection at
$\iM=20.2$ \citep{Richards04,VandenBerk04}.\footnote{This estimate of
completeness considers only sources with AGN-dominated optical/UV
spectra. Additional optically-unremarkable AGNs might also be missed.}
SDSS Data Release 2 (DR2) contains over 35,000 AGN spectra in \about
2630 deg$^2$ covering the observed 3800--9100\,\AA\ region
\citep{sdssDR2}. The initial sample selected for this work consists of
174 SDSS AGNs situated in areas covered by 49 medium-deep (11\,ks or
longer) \emph{ROSAT} PSPC pointings (see $\S$\ref{xray}).

RL AGNs tend to have higher X-ray luminosity for a given rest-frame UV
luminosity (i.e., flatter $\alpha$\si{ox} values) than RQ AGNs. It is
believed that the additional X-ray emission is associated with the
radio rather than the UV component \citep[e.g.,][]{XrayRadio}, so we
need to exclude the strongly RL objects if we want to study UV-X-ray
correlations and probe the energy generation mechanism intrinsic to
all AGNs. All but three of the 174 SDSS AGNs in the initial sample
have detections within 1.5$''$ or upper limits from FIRST.  Based on
the FIRST data and the \citet{Ivezic02} definition of radio-to-optical
monochromatic flux, we find nine strongly RL AGNs.  Following
\citet{Ivezic02}, we define the radio-loudness parameter, $R$, as the
logarithm of the ratio of the radio-to-optical monochromatic flux:
$R=\log(F\si{20\,cm}/F\si{\iM}) = 0.4(\iM-m\si{20\,cm})$, where
$m\si{20\,cm}$ is the radio AB magnitude \citep{ABmag},
$m\si{20\,cm}=-2.5\log(F\si{20\,cm}/3631\textrm{[Jy]})$, and \iM\ is
the SDSS $i$-band magnitude, corrected for Galactic extinction. We set
the radio-loudness threshold at $R=1.6$, excluding objects with
$R>1.6$.  Two of the remaining three AGNs with no FIRST coverage have
upper limits from the National Radio Astronomy Observatory Very Large
Array Sky Survey \citep[NVSS;][with typical sensitivity of
\about2.5\,mJy for 5$\sigma$ detections]{nvss} which are consistent
with our RQ definition. The radio loudness of the remaining AGN
(SDSS\,J2314$+$1407) is not tightly constrained by its NVSS limit
($R<1.8$). Taking into account that the NVSS constraint is close to
our chosen threshold of $R=1.6$ and that only \about 10\% of AGNs are
RL, it is unlikely that this single AGN is RL, so we retain it in the
main SDSS sample. Excluding the strongly RL AGNs reduces the sample of
174 to 165 objects.

The large optical wavelength coverage of the SDSS spectra allows
identification of BAL AGNs at $1.55<z<4.80$ via \ion{C}{4} absorption
(``High-ionization BALs'' -- ``HiBALs'') and $0.45<z<2.25$ via
\ion{Mg}{2} absorption (``Low-ionization BALs'' -- ``LoBALs''), as
well as weak-absorption AGNs \citep[i.e., absorption not meeting the
BAL criteria of][]{Weymann91}. BAL AGNs, with an observed fraction of
\hbox{10--15\%} in optically selected samples
\citep{Foltz90,Weymann91,Menou01,edrBAL,Hewett03,ReichardBAL}, are
known to be strongly absorbed in the soft X-ray band and thus to have
steep $\alpha$\si{ox} values \citep[e.g.,][]{BLW,Gallagher}. There are
20 AGNs with some UV absorption in the SDSS RQ AGN sample of 165, ten
of which are BAL AGNs by the traditional definition
\citep[troughs deeper than 10\% of the continuum, at least
2000\,km\,s$^{-1}$ away from the central emission wavelength, spanning
at least 2000\,km\,s$^{-1}$;][]{Weymann91}.  Eight of the BAL AGNs are
HiBALs (out of a possible 67 AGNs with $1.55<z<4.80$), and there are
two LoBALs (out of a possible 116 AGNs with $0.45<z<2.25$). Only three
of the ten BALs are serendipitously detected in deeper
\emph{XMM-Newton} exposures (one LoBAL with $\alpha\si{ox}=-1.6$ and two
HiBAL with $\alpha\si{ox}=-1.7$, see \S~\ref{xray}), the remaining
seven BALs have $\alpha\si{ox}$ upper limits ranging between $-1.4$
and $-2.0$, depending on the sensitivity of the \emph{ROSAT}
exposures. Exclusion of the 10 BAL AGNs reduces the sample from 165
to 155 objects. We expect there to be $\approx8$ more HiBAL and
$\lesssim1$ more LoBAL AGNs \citep[for a typical LoBAL:HiBAL ratio of
1:10;][]{ReichardBAL} which we are unable to identify because of a
lack of spectral coverage in the \ion{C}{4} or \ion{Mg}{2} regions. We
will estimate the effects of missed BALs on our sample correlations by
selectively excluding the steepest $\alpha\si{ox}$ sources in the
appropriate redshift intervals.

Three-quarters of the AGNs (117 objects) in the main SDSS sample of
155 allow direct measurement of the rest-frame 2500\,\AA\
monochromatic flux, $F_{\nu}$(2500\,\AA), from the SDSS spectrum. SDSS
DR2 reductions have substantially improved spectrophotometry relative
to earlier data releases (better than 10\% even at the shortest
wavelengths,\footnote{Details about the spectrophotometry can be found
at http://www.sdss.org/dr2/products/spectra/spectrophotometry.html.}
see also $\S$4.1 of Abazajian et al. 2004) but do not include corrections for
Galactic extinction. To correct the SDSS monochromatic flux
measurements for Galactic extinction we use the \citet{dustmaps} dust
infrared emission maps to estimate the reddening, $E(B-V)$, at each
AGN position\footnote{The code is available at
http://www.astro.princeton.edu/$\sim$schlegel/dust/index.html.} and
the \citet{Nandy75} extinction law with $R=A\si{V}/E(B-V)=3.14$ to
estimate the Galactic extinction, $A\si{V}$, as a function of
wavelength. The Galactic extinction correction is $<$10\% at
2500\,\AA\ in \about 80\% of the cases considered.

\begin{figure}
\epsscale{1.0}
\plotone{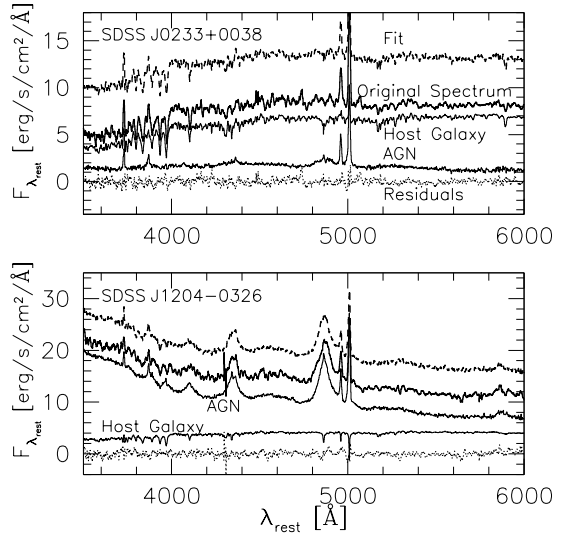}
\caption{Example high S/N AGN $+$ host-galaxy decompositions for two 
low-redshift spectra with dominant host-galaxy (upper) and AGN (lower)
contributions. The original spectrum is shown with the thick solid
line (smoothed to \about 11\,\AA\ resolution) and the eigenspectrum
fit with the thick dashed line (displaced by $+$5 monochromatic flux
units for clarity). The AGN and host-galaxy components are given
separately below with thin solid lines and the fit residuals are shown
with thin dotted lines in each case.
\label{specfits}} 
\end{figure}

The remaining quarter (38 objects) of the main SDSS sample AGNs lack
2500\,\AA\ rest-frame coverage in the observed 3800--9100\,\AA\ 
spectroscopic range. We use spectroscopic monochromatic flux
measurements at rest-frame 3700\,\AA\ (30 AGNs with $z<0.5$) and 
1470\,\AA\ (8 AGNs with $2.7 \leq z \leq 4.5$) with the appropriate
optical spectral slopes, $\alpha$\si{o} (assuming
$F_{\nu}\propto\nu^{\alpha\si{o}}$), to determine the monochromatic flux
at 2500\,\AA.  Based on over 11,000 AGNs from DR2 with both 1470\,\AA\ 
and 2500\,\AA\ monochromatic flux measurements, we estimate that an optical
slope of $\alpha\si{o}=-0.73$ gives the best agreement between the direct
2500\AA\ and $F_{\nu}$(1470\,\AA)-extrapolated monochromatic flux
measurements. This is redder (steeper) than the ``canonical'' AGN
slope over the optical-and-UV region of $\alpha\si{o}=-0.5$
\citep{Richstone80} because of the presence of the ``small blue bump''
\citep[see the discussions on the variation of spectral slope with the
rest-wavelength measurement range in][]{Natali98,Schneider01,DanSDSS}.
The error of the $F_{\nu}$(2500\,\AA) estimate due to the
$F_{\nu}$(1470\,\AA) extrapolation is typically less than 25\%.  A
canonical slope of $\alpha\si{o}=-0.5$ between 2500\,\AA\ and
3700\,\AA\ provides good agreement between the direct 2500\,\AA\ and
the $F_{\nu}$(3700\,\AA)-extrapolated monochromatic fluxes, based on
2,400 DR2 AGNs with $0.5<z<0.8$. The error in $F_{\nu}$(2500\,\AA)
expected due to variations in the 2500--3700\,\AA\ optical slope is
typically less than 20\%. In addition, because the direct
$F_{\nu}$(2500\,\AA) measurement includes a varying contribution from
\ion{Fe}{2} emission, $F_{\nu}$(2500\,\AA) could overestimate the
true nuclear flux by 10--25\% (as determined from \about 40
\ion{Fe}{2}-subtracted main-sample AGNs and comparison of
$F_{\nu}$(2500\,\AA) and the relatively \ion{Fe}{2}-free
$F_{\nu}$(2200\,\AA) measurement of \about 106 main sample AGNs),
leading to a $<3$\,\% error in $\alpha\si{ox}$. The possible
overestimate of $F_{\nu}$(2500\,\AA) due to \ion{Fe}{2} emission does
not correlate with luminosity or redshift and has no material effect
on the subsequent analysis.

An additional correction is necessary for the $F_{\nu}$(2500\,\AA)
estimates for low-redshift AGNs. If not subtracted, the host-galaxy
contributions of the 36 AGNs with $z<0.55$ could lead to potentially
large overestimates of rest-frame monochromatic UV fluxes of the AGNs. To
obtain a reliable estimate of the AGN contribution at 2500\,\AA\ for
the $z<0.55$ AGNs, we fit each observed spectrum with host-galaxy plus
AGN components. The host-galaxy and AGN components were created using
the first 3--20 galaxy and AGN eigenspectra obtained from large SDSS
samples by \citet{YipGal,YipQSO}. In the AGN (host-galaxy) case,
$>$90\% of the variation is explained by the first five (three)
eigenspectra. Two high S/N example fits are shown in
Figure~\ref{specfits}. The host-galaxy corrections (as measured at
3700\,\AA) are negligible for six of the 36 low-redshift AGNs and are
$\leq$20\% for 20 additional AGNs. 

\begin{figure}
\epsscale{1.0}
\plotone{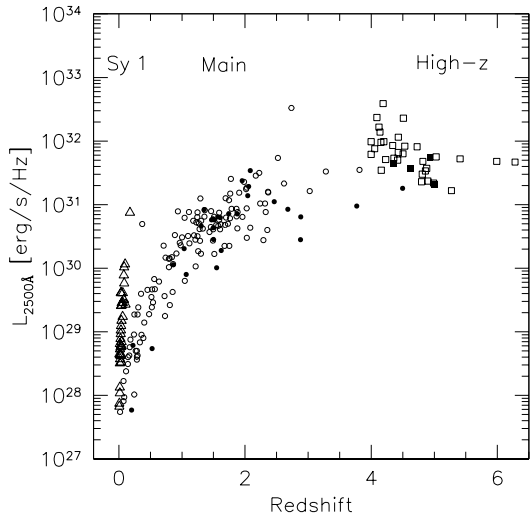}
\caption{The rest-frame monochromatic UV luminosities of the SDSS main
(circles), high-$z$ (squares), and Sy\,1 (triangles) samples
vs. redshift.  Open symbols indicate X-ray detections.
\label{l2500_z}} 
\end{figure}

In what follows, we use
$f\si{2500\,\AA}=\log({F_{\nu}(\textrm{2500\,\AA})}[\textrm{Jy}])$ to
denote the logarithm of the rest-frame monochromatic UV flux at
2500\,\AA, and
$l\si{uv}=\log({L_{\nu}(\textrm{2500\,\AA})}[\textrm{erg\,s$^{-1}$\,Hz$^{-1}$}])$
to denote the logarithm of the corresponding monochromatic
luminosity. The rest-frame monochromatic UV flux (no band-pass
correction was applied) and monochromatic luminosity (band-pass
corrected) measurements for the main SDSS sample of 155 objects are
presented in columns 7 and 12 of Table~\ref{tab1}, with the
spectroscopic redshift in column 2, the SDSS PSF \iM-band
extinction-corrected apparent magnitude in column 13, and the
$\Delta(g-i)$ color in column 14. The AGNs in Table~\ref{tab1} are
referenced by their unique SDSS position, J2000:
``SDSS\,JHHMMSS.ss$\pm$DDMMSS.s'', which will be shortened to
SDSS\,JHHMM$\pm$DDMM when identifying specific objects below.
Figure~\ref{l2500_z} presents the monochromatic luminosity at 2500\,\AA\
vs. redshift for the main SDSS as well as the high-$z$ and Sy\,1
samples. The selection bias toward more luminous AGNs at higher
redshift is evident.

\begin{figure}
\epsscale{1.0}
\plotone{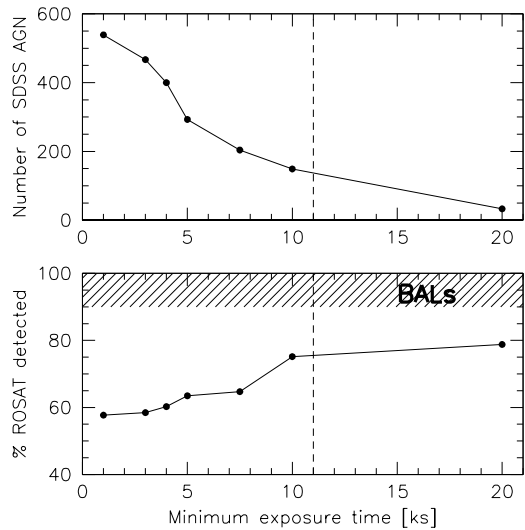}
\caption{Number of SDSS DR2 AGNs within the inner 19$'$ of \emph{ROSAT} 
PSPC fields (top) and percentage of SDSS AGNs detected by \emph{ROSAT}
(bottom) as a function of the minimum \emph{ROSAT} exposure time. No
RL or BAL AGNs were removed. The estimates were done at discrete
intervals given by the solid circles; the connecting lines are meant
to guide the eye. The vertical dashed lines show our choice of minimum
exposure time. Our final detection fraction (see Table~\ref{tab2}) is
$>$75\% for a minimum exposure time of 11\,ks due to the addition of
\emph{XMM-Newton} and \emph{Chandra} detections. The hatched region 
in the bottom plot is an approximate region taken by BALs in optical
surveys (see \S~\ref{optical}), bounding the realistically achievable
X-ray detection fraction in optical samples to a maximum of
\hbox{\about 85--90\%}.
\label{Texp_detected}} 
\end{figure}

\subsection{X-ray Detections}
\label{xray}

In order to ensure a high soft X-ray detection fraction for the
optically selected AGNs, we start with a subsample consisting of SDSS
AGNs falling within the inner 19$'$ of 49 \emph{ROSAT} PSPC
observations longer than 11\,ks. The median total exposure time is
\about 16.7\,ks with individual pointing exposure times ranging
between 11.8 and 65.6\,ks.\footnote{The effective exposure times for
  individual sources (given in Table~\ref{tab1}) will be shorter,
  depending on the source off-axis angle.} This approach does not
introduce biases into the main sample since the SDSS does not
specifically target \emph{ROSAT} pointed-observation areas, and we
exclude one SDSS AGN which was targeted as a \emph{ROSAT} source but
failed the SDSS AGN color selection. At the time of writing, the
completed \emph{ROSAT} mission has the advantage (compared to
\emph{Chandra} and \emph{XMM-Newton}) of a large-area, uniformly
reprocessed, and validated dataset. Figure~\ref{Texp_detected}
illustrates the trade-off between large sample size and high X-ray
detection fraction of SDSS AGNs in \emph{ROSAT} PSPC pointed
observations (no BALs or RL AGNs were removed for this plot).
Pointings with exposure times $\gtrsim10$\,ks are necessary to achieve
\hbox{70--80\%} detections in statistically large samples of SDSS DR2
AGNs. Note that detection fractions of 100\% are unrealistic to expect
with serendipitous, medium-deep, soft X-ray coverage of optical AGN
samples. For example, most BALs, comprising \hbox{10--15\%} of optical
samples, will remain X-ray undetected. In our initial sample, none of
the ten known BALs is detected with \emph{ROSAT}, and only three of the
ten are detected in deeper \emph{XMM-Newton} exposures. The highest
realistically achievable detection fraction for optical samples is
85--90\%, compared to 81\% in our main sample (see $\S$\ref{df}).
Using the full PSPC field instead of the inner 19$'$ would result in a
six-fold increase of the X-ray coverage area available for SDSS
matches, but with larger uncertainties in the measured fluxes and an
increased fraction of non-detections. The selected subsample contains
155 SDSS AGNs in 49 \emph{ROSAT} PSPC pointings. The total solid angle
covered by the inner 19$'$ of these 49 pointings is \about15 deg$^2$
(\about 0.57\% of the DR2 area covered by spectroscopy). To avoid
large uncertainties in the X-ray flux measurements due to uncertain
source counts, we have excluded two AGNs which are close to the much
brighter X-ray source NGC\,4073 (2$'$ and 4$'$ from the pointing
center). We also replaced the \emph{ROSAT} flux of SDSS\,J1331$-$0150
(which falls within the cluster Abell~1750), and those of
SDSS\,J1242$+$0229, SDSS\,J0942$+$4711, and SDSS\,J0943$+$4651 (which
had \hbox{2--3$\sigma$} detections in the \emph{ROSAT} \hbox{0.5--2\,keV}
band), with their \emph{XMM-Newton} detections.

\begin{figure}
\epsscale{1.0}
\plotone{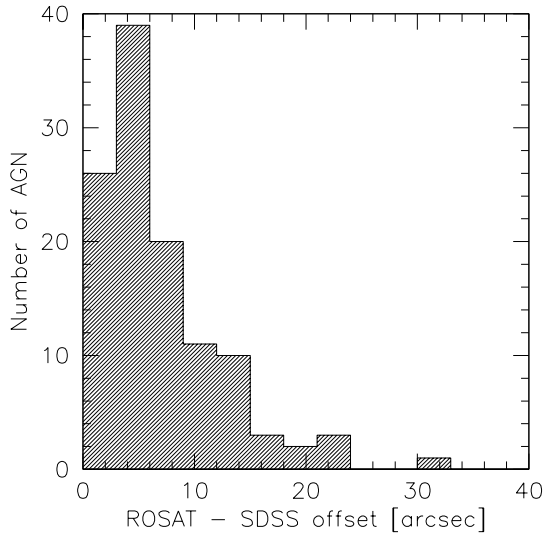}
\caption{Distribution of angular offsets between the SDSS and 
\emph{ROSAT} PSPC positions. The \emph{ROSAT} PSPC pixel size 
is 15$''$, and all matches are within \about 2 \emph{ROSAT}
pixels. The AGN with the largest (33$''$) offset is
SDSS\,J0255$-$0007, which was also detected as 1WGA J0255.1$-$0007
within 11$''$ of the SDSS position in the \hbox{0.24--2.0\,keV} 1\,WGA
catalog.
\label{offsets}} 
\end{figure}

We performed circular-aperture photometry using source photons with
energies of 0.5--2.0\,keV to obtain the count rates.  The exclusion of
$<0.5$\,keV photons was necessary to reduce the effects of absorption
due to neutral material (both in our Galaxy and intrinsic to the
AGNs), soft X-ray excesses, and \emph{ROSAT} PSPC calibration
uncertainties on the measured flux. The average aperture size used was
60$''$, with a range of 45$''$--90$''$ to accommodate the presence of
close companions and large off-axis angle sources. The count rates
were aperture corrected using the integrated \emph{ROSAT} PSPC
PSF.\footnote{http://wave.xray.mpe.mpg.de/exsas/users-guide/node136.html.}
The original apertures encircled $>$90\% of the \emph{ROSAT} flux in
\about 83\% of the cases; all aperture corrections were $<$20\% of the
measured count rate.  The background level was determined for each
field from a 14--25 times larger area with similar effective exposure
time to the source. The circular aperture for each source was centered
at the SDSS position in all but ten cases where the X-ray centroid in
an adaptively smoothed image\footnote{We use the Chandra Interactive
Analysis of Observations (CIAO) task csmooth,
http://cxc.harvard.edu/ciao3.0/ahelp/csmooth.html.} was \about 1-2
pixels (corresponding to \about 15--30$''$) away from the SDSS
position.  The distribution of SDSS--\emph{ROSAT} PSPC offsets (with
the X-ray centroids in the adaptively smoothed PSPC images serving as
\emph{ROSAT} positions) for the main sample is shown in
Figure~\ref{offsets}. The \about 33$''$ offset in Figure~\ref{offsets}
is that of SDSS\,J0255$-$0007, with an off-axis angle of 18$'$, which
is also detected in the \hbox{0.24--2.0\,keV} 1\,WGA
catalog\footnote{http://wgacat.gsfc.nasa.gov/wgacat/wgacat.html.}
\citep{WGA} with an X-ray flux consistent with our measurement, and a
positional offset of $11\pm50''$. In order to determine the number of
possible false SDSS--\emph{ROSAT} matches, we extract all unique
sources with off-axis angles \hbox{$<19'$} from the full \emph{ROSAT} PSPC
catalog obtained from the High Energy Astrophysics Science Archive
Research Center\footnote{http://heasarc.gsfc.nasa.gov/.} (HEASARC)
medium-deep \emph{ROSAT} pointings. To obtain the expected fraction of false
matches, we repeatedly shift all SDSS AGN positions by a random amount
in the range 0.1-1$^\circ$ and rematch them with the \emph{ROSAT} PSPC
catalog. The false-match fraction for SDSS--\emph{ROSAT} PSPC offsets
$<40''$ is $<0.1$\% (i.e., less than one source for the main SDSS
sample), which is further supported by our previous experience (see
VBS03).

\begin{figure}
\epsscale{1}
\plotone{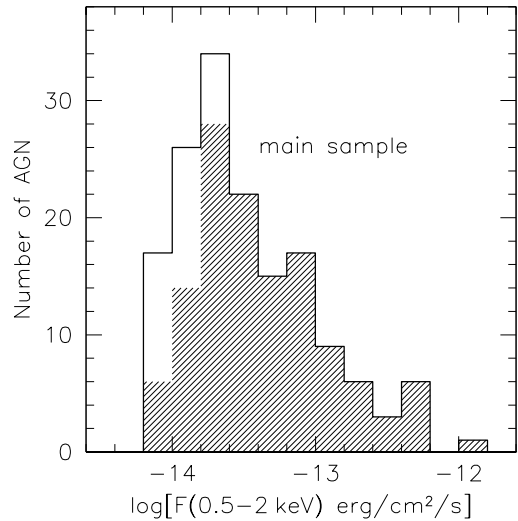}
\caption{Distribution of X-ray fluxes (hatched histogram) 
and upper limits (open histogram) for the SDSS main sample.
\label{XrayFlux}} 
\end{figure}

Table~\ref{tab1} gives the X-ray observation ID (column 3), effective
exposure time (4), off-axis angle (5), total source counts (6),
logarithm of the \hbox{0.5--2.0\,keV} flux (8), logarithm of the rest
frame 2\,keV monochromatic flux -- $f\si{2\,keV}$, not band-pass corrected
(9), and logarithm of the rest frame 2.0\,keV monochromatic luminosity --
$l_x$, band-pass corrected (11) for each source in the main SDSS
catalog. The
\hbox{0.5--2\,keV} flux histogram of the main SDSS sample is shown in
Figure~\ref{XrayFlux}. The soft X-ray detection limit for the inner
19$'$ of the medium-deep \emph{ROSAT} observations used here is \about
$2\times 10^{-14}$\,erg\,cm$^{-2}$\,s$^{-1}$.  The fluxes were
estimated using
PIMMS\footnote{http://heasarc.gsfc.nasa.gov/docs/software/tools/pimms\_install.html}
assuming a power-law X-ray spectrum with photon index $\Gamma=2$ and
the Galactic hydrogen column density obtained by
\citet{stark}. Previous studies suggest that AGN photon indices do not
vary systematically with redshift \citep[e.g.,][]{Page03,VignaliHighZ}, 
although the scatter (\about 0.5) around the mean value is substantial 
for all redshifts. Assuming a constant $\Gamma$, when in reality 
$1.5<\Gamma<2.5$ for the different sources, affects our flux measurements 
by $\lesssim4$\%. Four of the selected 155 SDSS AGNs were the targets 
of their respective \emph{ROSAT} pointed observations (marked by note 
1 in Table~\ref{tab1}). Their inclusion in the main sample could have a
small effect on the sample correlations, as the four AGNs do not
comply with our selection criteria -- optically selected AGNs
\emph{serendipitously} observed in medium-deep \emph{ROSAT} pointings. 
Three of the four AGNs are not substantially different in their rest-frame
UV and X-ray properties from the rest of the sample, while
SDSS\,J1701$+$6412 is the UV-brightest AGN in the main sample. We opt
to retain the four \emph{ROSAT} targets in the main sample, while
ensuring that their presence has no material effect on any of our
conclusions (see \S~\ref{luvlx} and \S~\ref{aox}).

A total of 40 of the 155 SDSS AGNs (26\%) were not detected in the
\hbox{0.5--2.0\,keV} band by \emph{ROSAT}. One of the 40 SDSS AGNs,
SDSS\,J1400$+$6225, is not detected by \emph{ROSAT} but is detected
serendipitously on CCD~S2 of a \emph{Chandra} ACIS-S \citep{AcisRef}
observation. We used ACIS Extract \citep{AEmanual}, which utilizes
\emph{Chandra} Interactive Analysis of Observations (CIAO
v.3.0.2)\footnote{http://cxc.harvard.edu/ciao/} tools, to estimate the
\hbox{0.5--2.0\,keV} flux. Nine additional AGNs with \emph{ROSAT} upper limits were
serendipitously detected in \emph{XMM-Newton} \citep{XmmRef}
observations, as indicated in column (3) of Table~\ref{tab1}. We use
the count rates in the \hbox{0.5--2\,keV} band of the first
\emph{XMM-Newton} serendipitous source catalog -- \hbox{1XMM
SSC}\footnote{http://xmmssc-www.star.le.ac.uk/newpages/xcat\_public.html}
\citep{xmmssc1}, whenever available (four sources), to obtain the \emph{XMM-Newton}
fluxes. For the remaining five \emph{XMM-Newton} detected sources which
are not in 1XMM~SSC, we use the source lists provided by the standard
\emph{XMM-Newton} processing to extract the
\hbox{0.5--2.0\,keV} count rates. When a source is detected by more 
than one \emph{XMM-Newton} European Photon Imaging Camera (EPIC)
instrument \citep{epicRef}, we average the estimated fluxes weighting
by the quoted errors and report the MOS total counts and effective
exposure times in Table~\ref{tab1}. An additional 14 sources detected
by \emph{ROSAT} are also detected by \emph{XMM-Newton}. The
\hbox{0.5--2.0\,keV} fluxes of these 14 AGNs agree within 0.4 dex 
(a factor of ~2.5) in 12 of the cases, and the \emph{XMM-Newton}
detections are more likely to be brighter by \about 30\%. Taking into
account that four of the \emph{ROSAT} detections are
\hbox{2--3$\sigma$} and that AGNs are variable on scales of hours to
years (see the discussion of AGN X-ray variability in
$\S$~\ref{dispersions}), we consider this agreement adequate for
inclusion of the \emph{XMM-Newton} detected AGNs with no \emph{ROSAT}
detections into our sample.

A total of 14 AGNs in our main sample have \emph{XMM-Newton} (13/14)
or \emph{Chandra} (1/14) detections replacing the \emph{ROSAT} upper
limits (10/14) or low-confidence/cluster-contaminated detections
(4/14).  The \emph{XMM-Newton}/\emph{Chandra} observations could be
more likely to ``catch'' the SDSS AGNs in a high-luminosity state, if
the difference between the \emph{ROSAT} limiting flux and the
\emph{XMM-Newton}/\emph{Chandra} detection flux is sufficiently small
in comparison to AGN variability. Four of the 14 AGNs with
\emph{XMM-Newton}/\emph{Chandra} detections have fluxes above their 
\emph{ROSAT} limits (\about 30\% higher) and could have been detected 
with \emph{XMM-Newton}/\emph{Chandra} only because they were in a
high-luminosity state. The remaining ten AGNs were detected in more-sensitive
\emph{XMM-Newton}/\emph{Chandra} observations.  On account of these
possible ``high-state'' detections and the tendency of some
\emph{XMM-Newton} detections to provide brighter \hbox{0.5--2.0\,keV}
fluxes than the corresponding \emph{ROSAT} detections, we will
consider the effect of excluding all 14
\emph{XMM-Newton}/\emph{Chandra} detected AGNs on the subsequent
correlations.

\subsection{The High-Redshift Sample}

To increase the redshift and luminosity coverage of the optically
selected AGN sample, we add an auxiliary sample of 36 AGNs at $z>4$.
These high-$z$ AGNs were selected from 44 AGNs specifically targeted
for X-ray imaging with \emph{Chandra} \citep[19 SDSS AGNs, 16 Palomar
Digital Sky Survey AGNs; Djorgovski et al.~1998; and seven AGNs from
the Automatic Plate Measuring facility survey,][]{APMagn} and
\emph{XMM-Newton} (2 SDSS AGNs) reported in Tables~3~and~A1 of
\citet{VignaliHighZ}. The 36 high-$z$ AGNs were selected from the
original 44 AGNs by excluding three strongly radio-loud ($R>1.6$) AGNs
and five BAL AGNs.  This sample is somewhat more heterogeneous in its
optical selection (although all $z<5.4$ high-$z$ AGNs would have made
the SDSS AGN target selection), contains only the highest rest-frame
UV luminosity AGNs, and was specifically targeted for X-ray
observations.  Consequently we carefully consider the effect of its
addition to the main sample on the \hbox{rest-frame UV-X-ray}
relations reported below.

\subsection{The Seyfert 1 Sample}

As noted in \S~\ref{intro}, the significance of UV-X-ray correlations
depends on the range of luminosities probed for each redshift.  The
SDSS selects photometric targets for spectroscopic follow-up in two
magnitude ranges -- low-redshift targets are magnitude limited at
$\iM<19.1$ and high-redshift targets at $\iM<20.2$. Mainly due to the
large solid angle covered by the SDSS, but also on account of its two
different optical flux limits, the main SDSS sample probes a luminosity range
of at least an order of magnitude at each redshift, except
at $z\lesssim0.2$ and $z\gtrsim3$. In order to increase the luminosity
range for low-redshift AGNs, we consider an additional sample of
Seyfert 1 galaxies with measurements from both
\emph{IUE} and \emph{ROSAT}. The majority of objects were selected
from the Seyfert~1 list of \citet{WalterB} to have direct monochromatic flux
measurements at both 2675\,\AA\ and 2\,keV (see their Table 1) and
$L\si{$\nu$}(\textrm{2500\,\AA})>10^{27.5}$\,erg\,s$^{-1}$\,Hz$^{-1}$. NGC
3516 \citep{Kolman93} was added to the \citet{WalterB} Seyfert~1 list,
and the I\,Zw~1 measurements were replaced with recent, more accurate
estimates from \citet{Gallo}. The monochromatic flux measurements at
2675\,\AA\ were not corrected for host-galaxy contamination, which we
expect to be small at this wavelength for most sources. We inspected
visually a few high S/N \emph{IUE} spectra which showed no strong
host-galaxy features. To exclude strongly radio-loud objects we
consider only Seyfert 1s with
$L\si{$\nu$}(\textrm{5\,GHz})<10^{25}$\,W\,Hz$^{-1}$, where we use the
5\,GHz fluxes from \citet{WalterB} and additional 1.4\,GHz flux
measurements (extrapolated to 5\,GHz) from
NED\footnote{http://nedwww.ipac.caltech.edu/}, and exclude all
Seyferts with unknown radio flux from FIRST or NVSS. Our final
Seyfert~1 list consists of 37 AGNs. This sample is not biased in the
sense that it includes only X-ray detections of known optical AGNs. It is
not, however, purely optically selected; consequently, we evaluate all
correlations with and without the Sy\,1 subsample, to control for any
possible systematics.

\section{Correlation Analysis}

\subsection{Detection Fractions}
\label{df}

A high X-ray detection fraction, which minimizes the effects of
systematic pattern censoring and statistical assumptions, is
essential for accurate determination of AGN UV-X-ray properties (see
\S~\ref{intro}).  As can be seen in Figure~\ref{XrayFlux}, most
sources with
$F$\si{(0.5-2\,keV)}$>2\times10^{-14}$\,erg\,cm$^{-2}$\,s$^{-1}$ are
detected for an overall detection fraction of 126/155 (81\%) in the
main SDSS sample.  The X-ray detection fractions (X-ray detected
vs. total number) for the main, high-$z$ and combined (main, high-$z$,
and Sy\,1) samples are given in Table~\ref{tab2}.

\begin{figure}
\epsscale{1.0}
\plotone{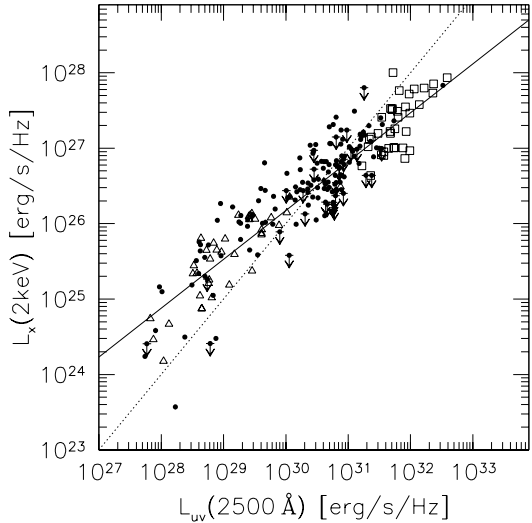}
\caption{The 2\,keV vs. 2500\,\AA\ monochromatic luminosities for the SDSS main 
sample (solid circles), the high-$z$ sample (open squares), and the
Sy\,1 sample (open triangles). Arrows indicate upper limits. The solid
line is the best-fit linear relation for the combined sample:
$l\si{x}=0.648l\si{uv} + 6.734$, and the dotted line is an
arbitrarily normalized line with a slope of one.
\label{luv_lx}} 
\end{figure}

\subsection{Monochromatic Optical/UV and X-ray Luminosities}
\label{luvlx}

Figure~\ref{luv_lx} shows the relation between the 2\,keV and
2500\,\AA\ monochromatic luminosities. The correlation is significant
at the 11.5$\sigma$ (7.4$\sigma$) level, after the redshift dependence
of both quantities and all upper limits are taken into account for the
combined sample of 228 AGNs (the main SDSS sample of 155 AGNs). The
partial Kendall's correlation coefficient \citep{AS96} is
$\tau$\si{12,3}=0.38 ($\tau$\si{12,3}=0.28) for the combined (main)
sample (see Table~\ref{tab3}).  

In order to test the partial-correlation method, we created mock
datasets with variable dispersion and strong redshift dependence. We
consider cases of (1) no relation between the dependent and
independent variables and (2) a linear relation between the dependent
and independent variables. In both cases we assume that the UV
monochromatic luminosity is a polynomial function of redshift with a
luminosity range of about an order of magnitude at each redshift,
which includes a normally distributed dispersion (with standard
deviation equal to the observed regression residuals from
Table~\ref{tab3}) to both the UV and X-ray monochromatic luminosities
to simulate the uncertainty due to variability and measurement
errors. The ``true relation'' simulation further assumes that
Eqns.~1--3 given below hold, while the ``no relation'' simulation
assumes that the X-ray monochromatic luminosity is a different
polynomial function of redshift. When we match the observed redshift
distribution and number of X-ray upper limits, we confirm the
existence of the mock-linear relations with similar statistical
significance to the significance found for the real datasets, \about
12--14$\sigma$ in the mock-combined and \about 8--10$\sigma$ in the
mock-main simulated samples, weakly dependent on the ratio (varying
between 0.5 and 2.0 in our simulations) of dispersions assumed for the
dependent and independent variables.

For ``no relation'' simulations, spurious correlations of up to
4$\sigma$ in the mock-main and up to 7$\sigma$ in the mock-combined
sample are possible. The apparent high significance of the
``no-relation'' simulations is caused by our lack of knowledge of the
true mean dependence of the monochromatic luminosity on redshift in
the UV and X-ray bands separately, combined with the observational
constraint on the range of luminosities probed at each redshift. The
simulation set-up is further affected by the fact the observations
constrain only the total dispersions along the
\hbox{$l\si{x}$--$l\si{uv}$}, \hbox{$l\si{uv}$--$z$}, and 
\hbox{$l\si{x}$--$z$} relations, without strong constraints on the 
contribution of variability and measurement error. By necessity, the
polynomial fits we use in the simulation to represent the mean
\hbox{$l\si{uv}$--$z$} and \hbox{$l\si{x}$--$z$} relations are 
very similar, and consequently simulations with significant spurious
correlations are possible. However, in no simulation where we match
the observed \hbox{$l\si{uv}$--$z$} and \hbox{$l\si{x}$--$z$}
distributions (in both their mean relations and dispersions) as well
as the observed \hbox{$l\si{x}$--$l\si{uv}$} dispersion, are the ``no
relation'' correlations found significant enough to cause the observed
\hbox{$l\si{x}$--$l\si{uv}$} correlations.  Additionally, in all
simulated cases the significance of the ``true relation'' simulation
is sufficiently higher than the corresponding ``no relation'' case,
allowing for easy distinction between the two.  Consequently, we are
convinced that 11.5$\sigma$ (7.4$\sigma$) level correlations found for
the combined (main) samples are unlikely to arise on account of the
strong redshift dependence of the UV and X-ray monochromatic
luminosities.

The best-fit relations, assuming no redshift dependence (see the
discussion below), are
\begin{eqnarray}
l\si{x}(l\si{uv})=(0.645\pm0.034)l\si{uv}+(6.851\pm1.036) \nonumber \\
\textrm{main sample} \\
l\si{x}(l\si{uv})=(0.639\pm0.026)l\si{uv}+(7.026\pm0.804) \nonumber \\
\textrm{main$+$high-$z$} \\
l\si{x}(l\si{uv})=(0.648\pm0.021)l\si{uv}+(6.734\pm0.643) \nonumber \\
\textrm{main$+$high-$z$$+$Sy\,1} 
\end{eqnarray}
(the excess precision quoted is useful for plotting purposes). In all
cases the fits given above were obtained using the EM algorithm for
censored data from ASURV; the Buckley-James method from ASURV returns
results consistent within $1\sigma$. The resulting slope is less than
one in all cases, implying a changing ratio between the 2\,keV and
2500\,\AA\ monochromatic luminosities with rest-frame UV luminosity. The
residual scatters around the linear relationships are 0.39, 0.37, and
0.36 (in $\log$ units) for the main, main$+$high-$z$, and combined
samples, respectively (see Table~\ref{tab3}). Removing the four AGNs
which were targets of \emph{ROSAT} pointings or the 14 AGNs with
\emph{XMM-Newton}/\emph{Chandra} X-ray photometry, has no material
effect on the parameters of the linear regression and only slightly
decreases the significance of the correlation (on account of the
decrease in sample size and the consequent slight increase in the
fraction of upper limits when $\gtrsim10$ detections are excluded). In
order to check for any effect of the unidentified HiBALs/LoBALs remaining in
our sample, we exclude the 9 steepest $\alpha\si{ox}$ sources with
$z<1.55$ from the main sample before performing the correlation. The
linear regression parameters for the main sample remain unchanged
within the quoted errors, with
\hbox{$l\si{x}=(0.65\pm0.03)l\si{uv}+(6.8\pm1.0)$}.  Similarly for the
combined sample, assuming a 10\% observed HiBAL fraction and taking
into account that there are 146 AGNs without
\ion{C}{4} coverage, we exclude the 15 steepest $\alpha\si{ox}$ 
sources with $z<1.55$ or $z>4.8$, before repeating the correlation
analysis. We find \hbox{$l\si{x}=(0.63\pm0.02)l\si{uv}+(7.3\pm0.6)$},
consistent within 1$\sigma$ with Eqn.~3 above. Removing an additional
10 SDSS AGNs from the main sample with some UV absorption which do not
satisfy the BAL criteria (see \S~\ref{optical}) also has no effect on
the correlation parameters, yielding
\hbox{$l\si{x}=(0.65\pm0.04)l\si{uv}+(6.7\pm1.1)$}. Constraining the
linear regression to the 81 AGNs with $1.55<z<4.80$, where BAL AGNs
are easy to exclude using the absorption blueward of \ion{C}{4}, we
obtain a slightly shallower slope for the
\hbox{$l\si{x}$--$l\si{uv}$} correlation, 
\hbox{$l\si{x}=(0.58\pm0.06)l\si{uv}+(8.8\pm1.8)$}, consistent with
Eqns.~1--3 within $1\sigma$. 

\begin{figure}
\epsscale{1.0}
\plotone{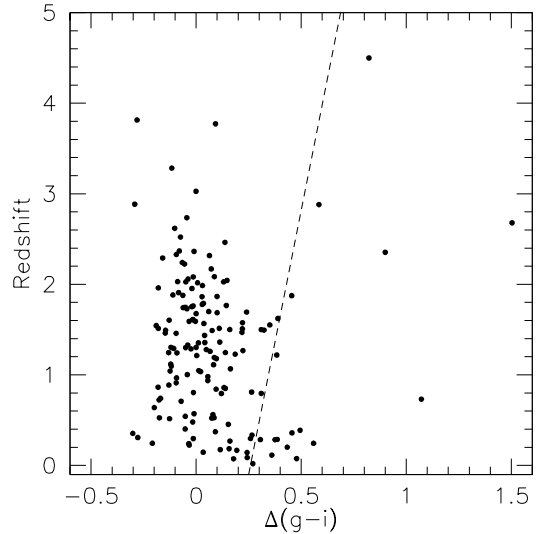}
\caption{Positions of the main-sample AGNs in the relative \gM$-$\iM\ 
color, $\Delta(g-i)$ vs. redshift diagram. The dashed line represents
SMC-type reddening as a function of redshift with $E(B-V)=0.04$
shifted redward by 0.2 to satisfy the \citet{Richards03}
dust-reddening definition. AGNs to the right of the dashed line can be
considered to be dust reddened \citep[see Figure~6
of][]{Richards03}. Note that the Lyman limit affects the \gM\
magnitudes of the four AGNs with $z>3.1$, rendering their relative
colors unreliable.
\label{rel_gr}}
\end{figure}

In order to probe the effects of any dust absorption in the rest-frame
UV on the \hbox{$l\si{x}$--$l\si{uv}$} relation, we use the relative
\gM$-$\iM\ AGN color, $\Delta(g-i)$. \citet{Richards03} have shown that
a $\Delta(g-i)$ vs. $z$ diagram, like the one presented in
Figure~\ref{rel_gr}, can be used to define a dust reddened AGN
subsample (to the right of the dashed line, see their
Figure~6). Excluding the 17 AGNs considered dust-reddened according to
\citet{Richards03} definition has no effect on the parameters of the
\hbox{$l\si{x}$--$l\si{uv}$} correlation in the main sample,
\hbox{$l\si{x}=(0.65\pm0.04)l\si{uv}+(6.8\pm1.1)$}.

\subsection{$\alpha$\si{ox} -- primary dependence on luminosity rather than redshift}
\label{aox}

\begin{figure}
\epsscale{1.0}
\plotone{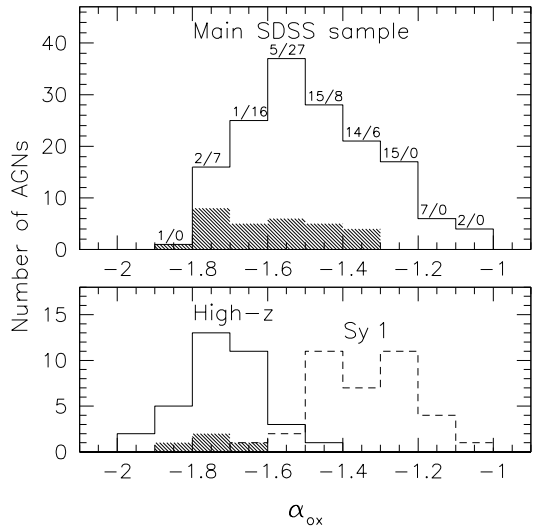}
\caption{$\alpha$\si{ox} distributions for the main SDSS (top), 
high-$z$ (bottom, solid line), and Sy\,1 (bottom, dashed line)
samples. The hatched histograms indicate X-ray upper limits. The two
numbers on the top of each bin in the top histogram give the numbers
of AGNs with $l\si{uv}<30.5$ (left) and $l\si{uv}>30.5$ (right) in
that bin.
\label{a_oxHist}} 
\end{figure}

Distributions of $\alpha$\si{ox} are presented in
Figure~\ref{a_oxHist} for the main SDSS (top) and the high-$z$ and
Sy\,1 (bottom) samples. The main SDSS sample has a median
$\alpha\si{ox}=-1.51$, compared to $\alpha\si{ox}=-1.72$ for the
high-$z$ sample and $\alpha\si{ox}=-1.34$ for the Sy\,1 sample. In
addition, as can be seen from the numbers on the top of each bin in
the top histogram of Figure~\ref{a_oxHist}, lower monochromatic luminosity
AGNs ($l\si{uv}<30.5$, left number) have flatter $\alpha\si{ox}$
indices compared to higher monochromatic luminosity AGNs ($l\si{uv}>30.5$,
right number).  It is therefore apparent that $\alpha\si{ox}$ is
correlated with rest-frame monochromatic UV luminosity and/or redshift. We
will show below that the primary dependence of $\alpha$\si{ox} is on
rest-frame monochromatic UV luminosity, while the redshift dependence is
insignificant.

\begin{figure}
\epsscale{2.0}
\plottwo{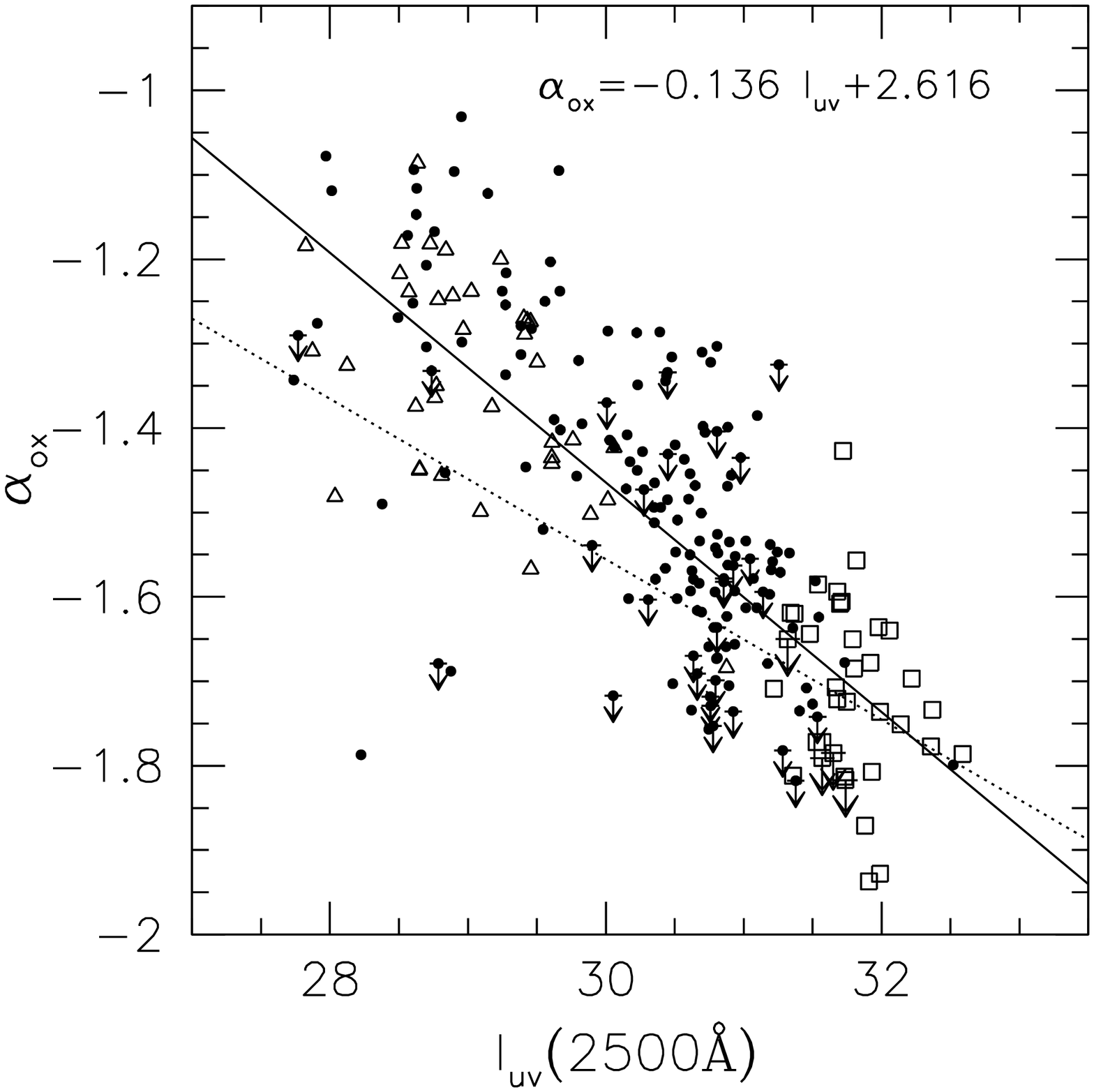}{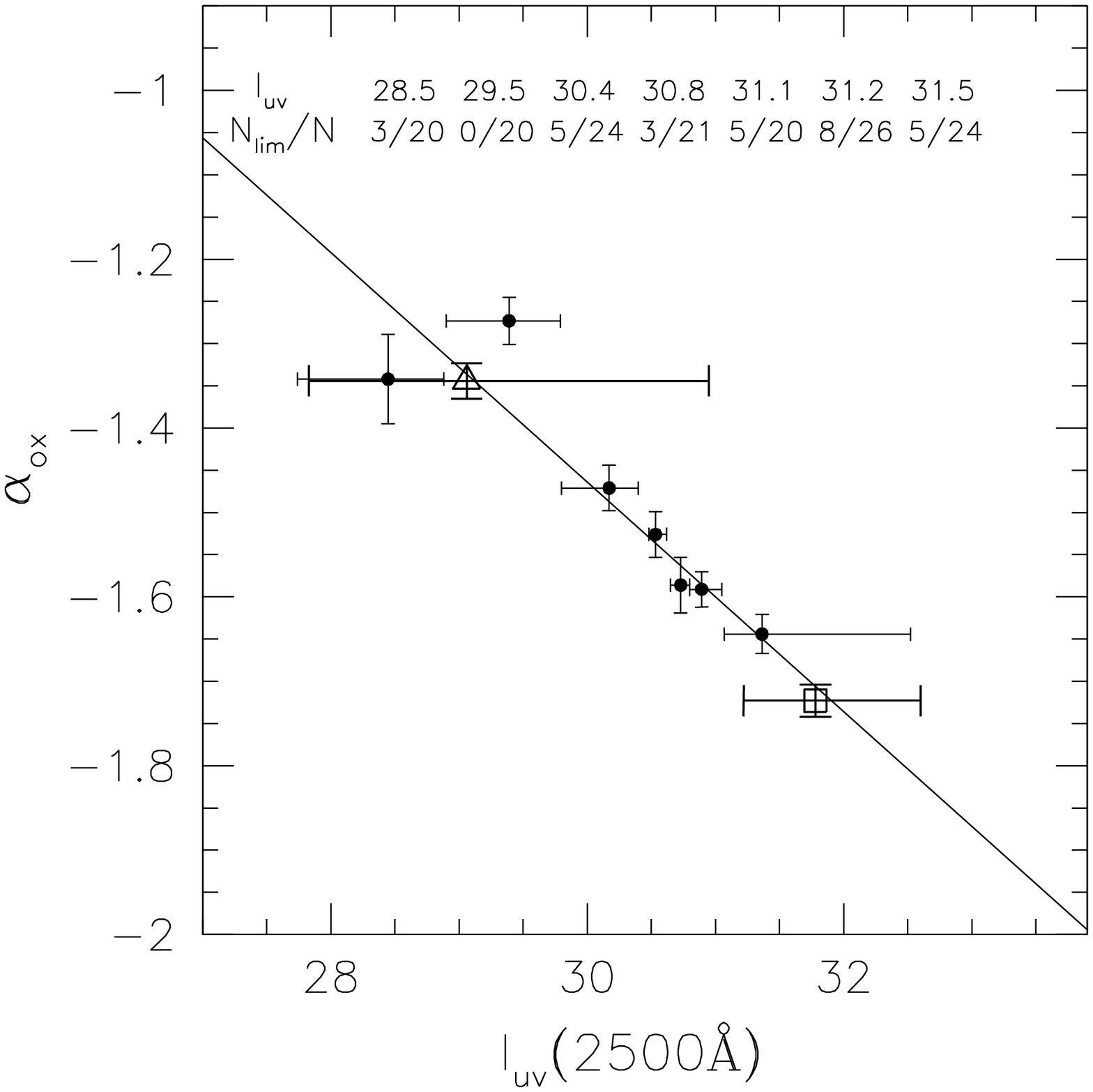}
\caption{\emph{Top:} $\alpha$\si{ox} dependence on the 2500\,\AA\
monochromatic luminosity. The main sample is given with solid circles, the
high-$z$ sample with open squares, and the Sy\,1 sample with open
triangles; arrows in the top panel indicate $2\,$keV upper
limits. The solid line is the linear relation found for the combined
sample (Eqn.~6), and the dotted line is the corresponding relation
from \citet[their Eqn.~4]{VignaliHighZ}. The anti-correlation is
significant at the \hbox{10.6$\sigma$} level for the combined
sample. \emph{Bottom:} Kaplan-Meier estimator of the mean
$\alpha$\si{ox} as a function of $l$\si{uv}. The numbers at the top
indicate the centers of the $l$\si{uv} bins and the number of limits
vs. the total number of AGNs in each bin.
\label{a_ox_luv}} 
\end{figure}

\begin{figure}
\epsscale{2.0}
\plottwo{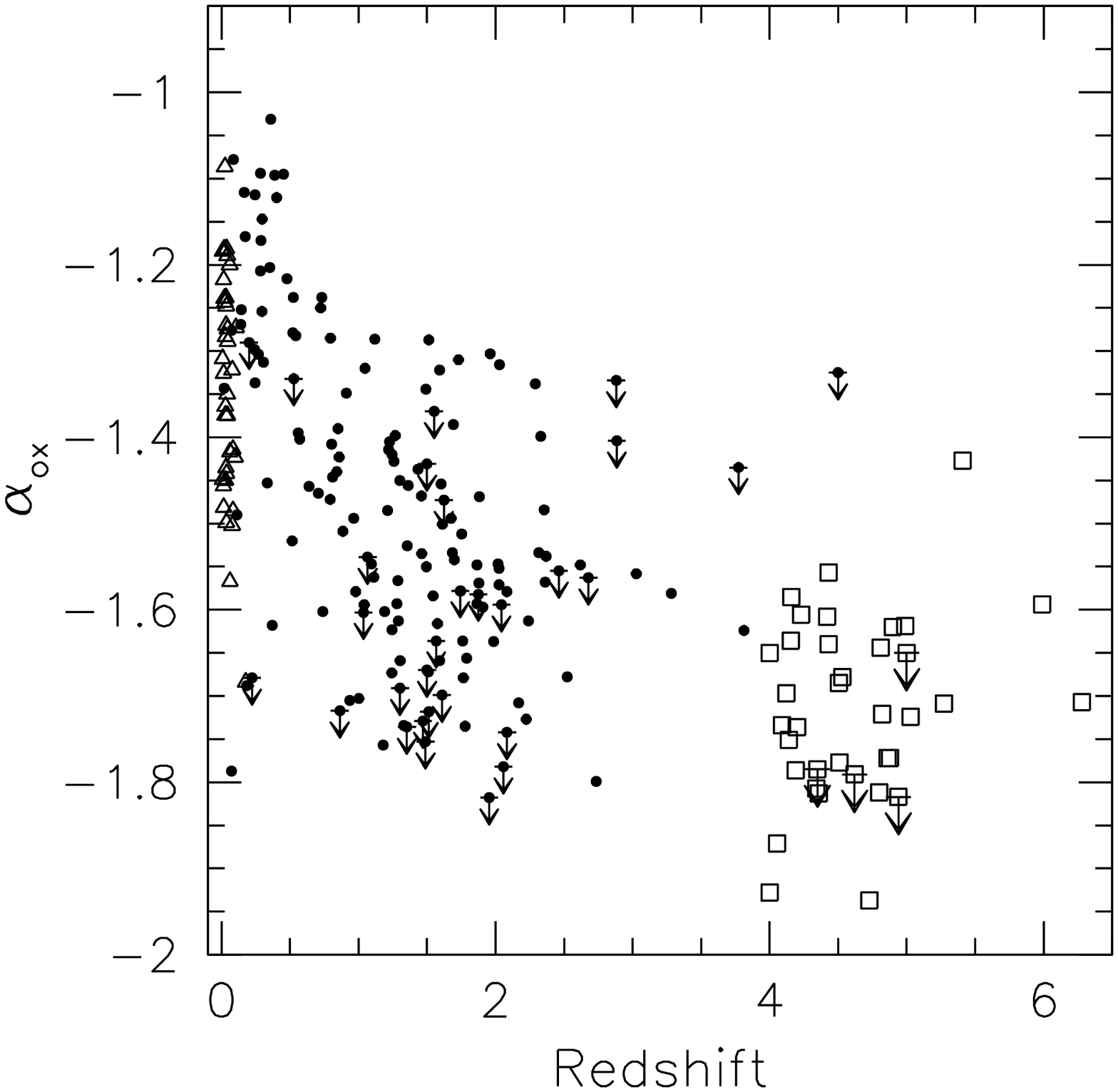}{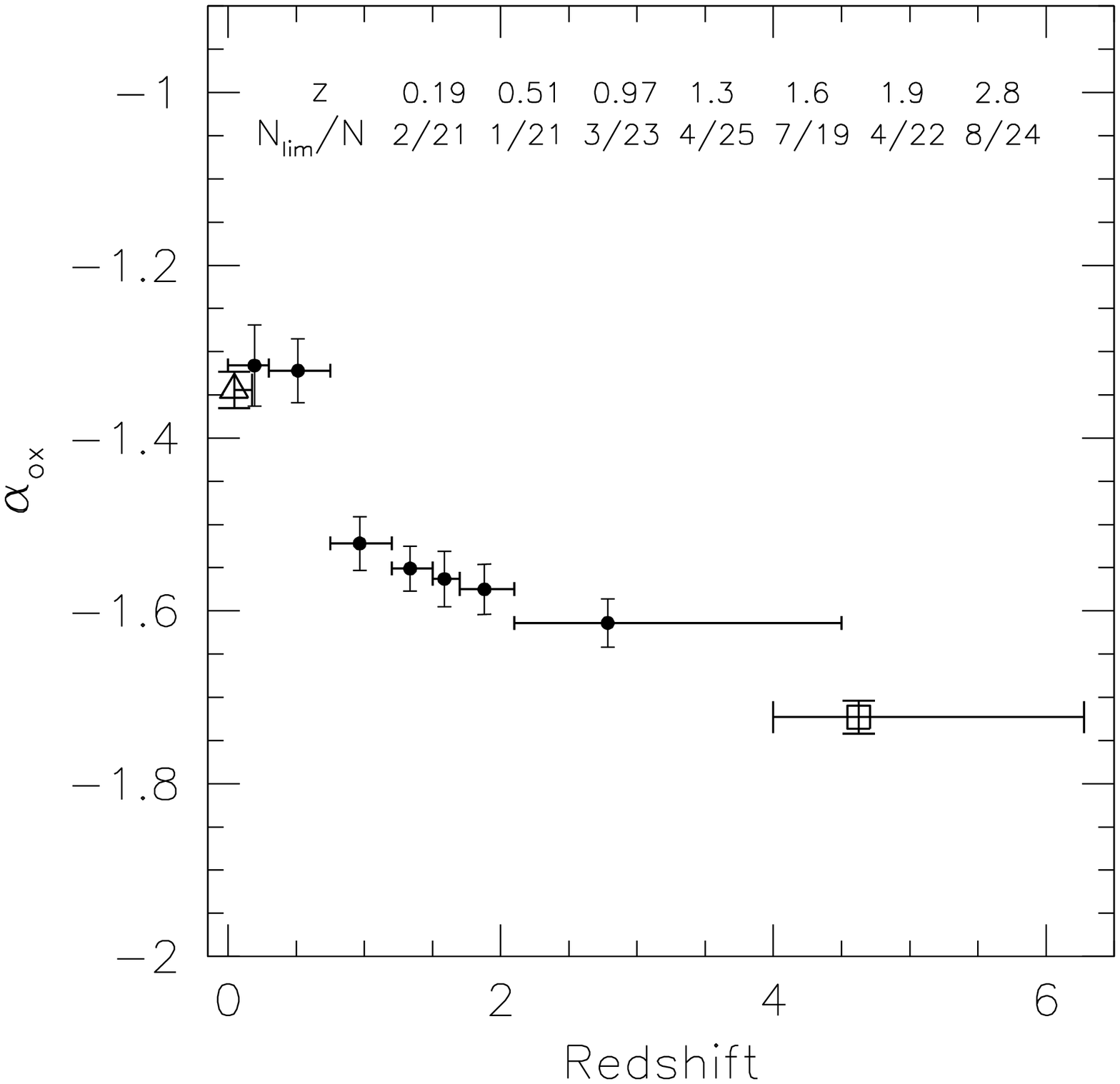}
\caption{ \emph{Top:} The correlation of $\alpha$\si{ox} with $z$ is only
\about 1$\sigma$ significant if the $l$\si{uv} dependence is taken 
into account (see \S~\ref{aox}). \emph{Bottom:} Kaplan-Meier estimator
of the mean $\alpha$\si{ox} as a function of redshift. Symbols and
numbers are as in Figure~\ref{a_ox_luv}.
\label{a_ox_z}} 
\end{figure}

\begin{figure}
\epsscale{2.0}
\plottwo{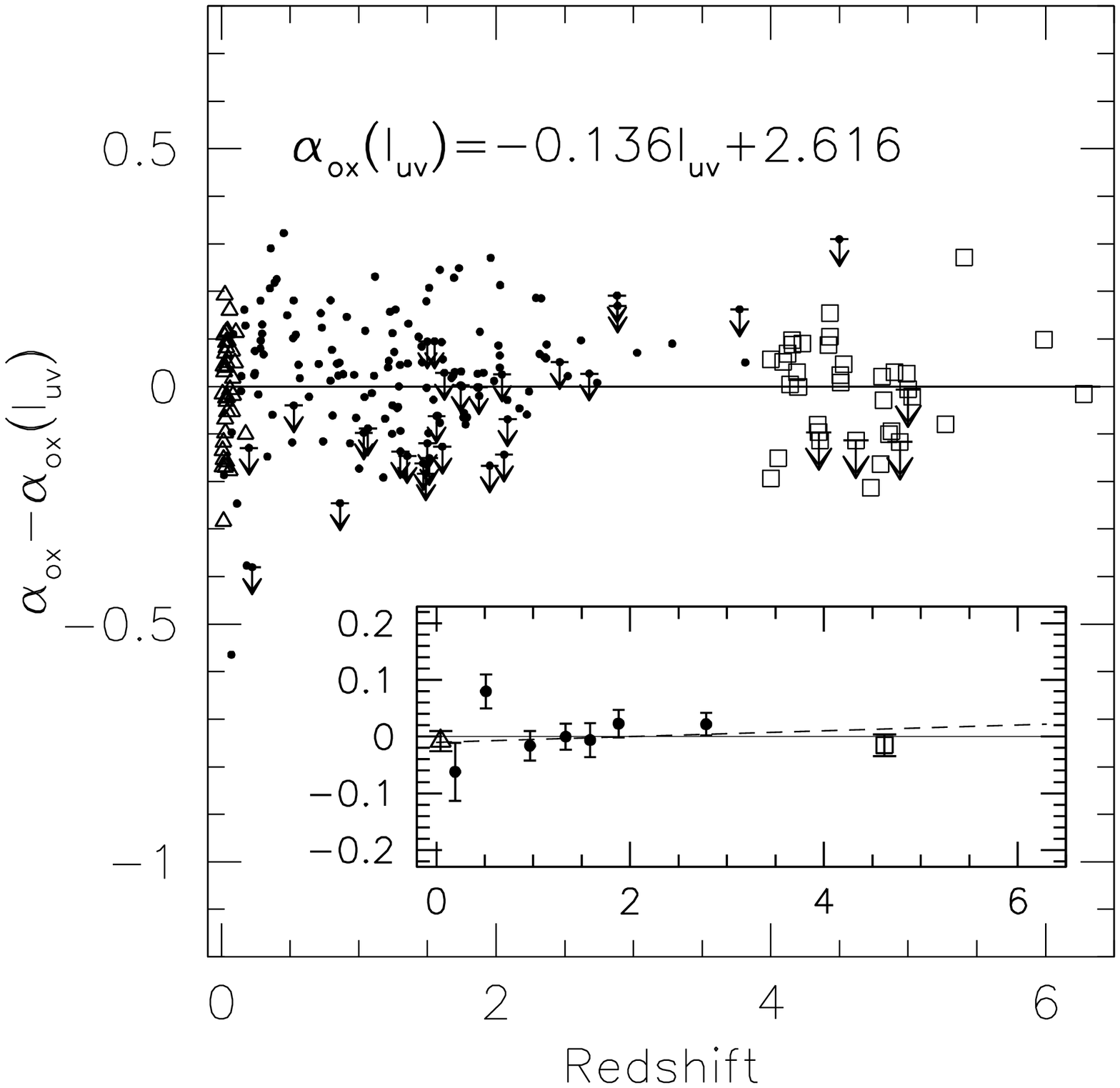}{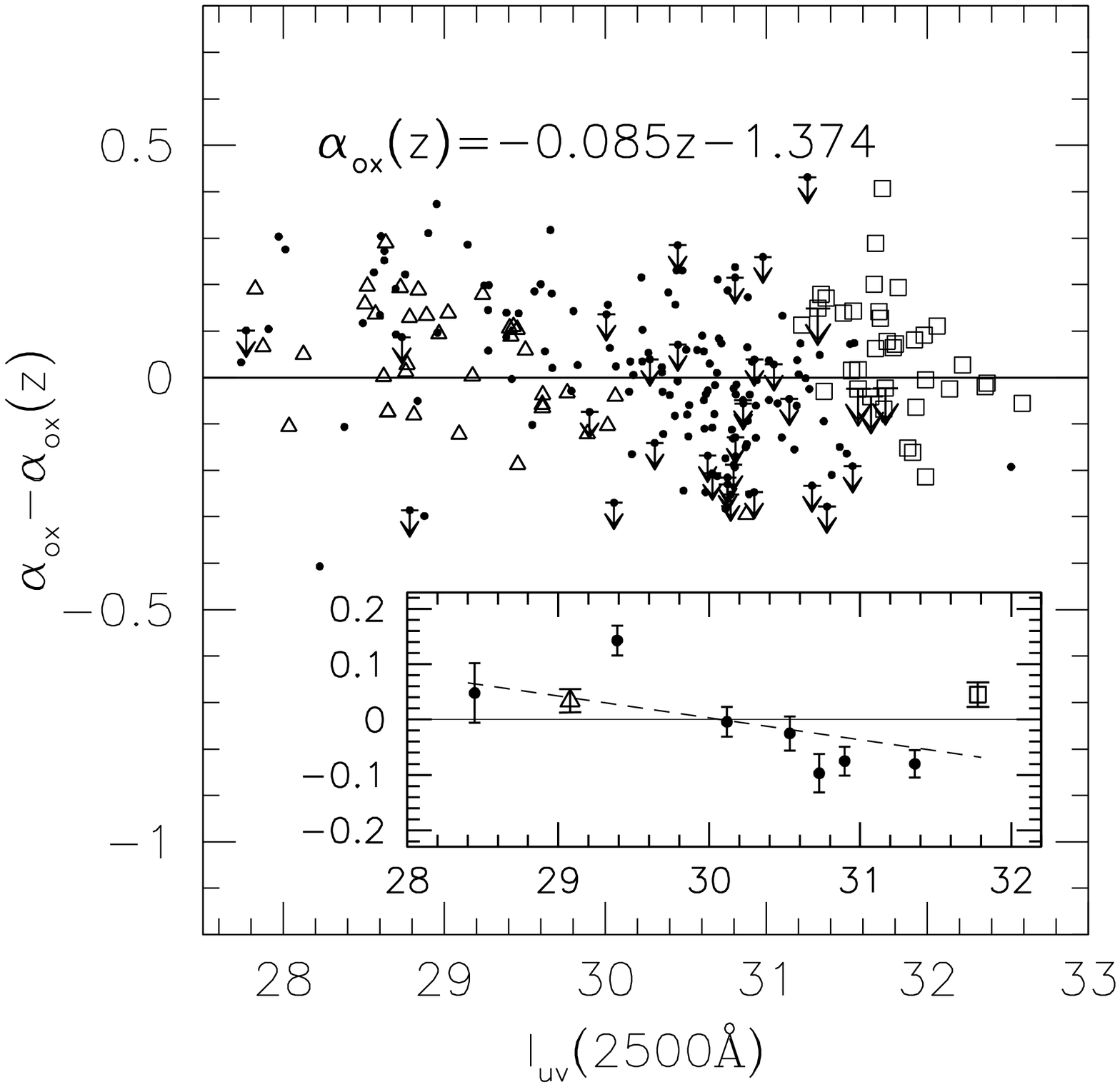}
\caption{Single-parameter fit residuals for an $\alpha$\si{ox} dependence 
on $l\si{uv}$ (top) or $z$ (bottom) for the combined sample. The
systematic residuals in the right plot support the idea that redshift
alone cannot be responsible for the observed variation in
$\alpha$\si{ox}. Symbols are as in Figure~\ref{a_ox_luv}. The small
inset plots in each panel give the Kaplan-Meier estimators of the
means of the $\alpha\si{ox}-\alpha\si{ox}(l\si{uv})$ (top) and
$\alpha\si{ox}-\alpha\si{ox}(z)$ (bottom) residuals in the same
redshift and monochromatic luminosity bins as in
Figures~\ref{a_ox_z}~and~\ref{a_ox_luv}, respectively. The dashed
lines in each inset plot correspond to the weighted linear regression
fits,
\hbox{$\left<\alpha\si{ox}-\alpha\si{ox}(l\si{uv})\right>=(0.005\pm0.012)z+(-0.010\pm0.023)$} (top) and
\hbox{$\left<\alpha\si{ox}-\alpha\si{ox}(z)\right>=(-0.039\pm0.020)l_{uv}+(1.184\pm0.603)$} (bottom).
\label{luv_z}}
\end{figure}

Figures~\ref{a_ox_luv}~and~\ref{a_ox_z} present the $\alpha$\si{ox}
dependence on $l$\si{uv} and redshift.\footnote{The rank correlation
analysis used in this paper is more general than linear correlation
methods. The ``rank coefficient'' is constructed by comparison of all
possible pairs of points, considering their relative positions rather
than exact values. Consequently the correlation results are unaffected
by the choice of $z$ instead of $\log(1+z)$ as the independent
variable.}  The optical/UV-to-X-ray index $\alpha$\si{ox} depends
primarily on $l$\si{uv} with a linear partial correlation coefficient
of $\tau\si{12,3}=-0.33$ ($\tau\si{12,3}=-0.30$) at a significance
level of 10.6$\sigma$ (7.4$\sigma$) for the combined (main) sample.
Table~\ref{tab3} presents the partial correlation statistics for
various AGN subsamples.  Taking into account that the
\hbox{$\alpha$\si{ox}--$z$} correlation coefficient changes from
negative (main and main$+$high-$z$ samples) to positive (combined
sample), and that the correlation significance level is always
$<1.1\sigma$, our Monte Carlo simulations suggest that any apparent
correlation could arise by chance due to the third variable
($l$\si{uv}) dependence.  To illustrate this using the combined
sample, we show in Figure~\ref{luv_z} the residuals for an assumed
$\alpha$\si{ox} dependence on a single parameter -- $l\si{uv}$ in the
top panel and $z$ in the bottom.  The structure of the residuals
confirms that an $\alpha$\si{ox} dependence on $l\si{uv}$ only is
adequate to describe the observed variation in $\alpha$\si{ox}, while
a redshift dependence alone is inadequate (as shown by the systematic
residuals). In fact, if we attempt to fit a relation of the form,
$\alpha\si{ox}=Al\si{uv}+Bz+C$ to the combined sample, the result is a
fit with $B$ equal to zero within the errors.  The linear regression
fits, taking into account the $\alpha$\si{ox} upper limits and
ignoring any redshift dependence, are
\begin{eqnarray}
\alpha\si{ox}(l\si{uv})=-(0.136\pm0.013) l\si{uv}+(2.630\pm0.398) \nonumber \\
\textrm{main sample}\\
\alpha\si{ox}(l\si{uv})=-(0.139\pm0.010) l\si{uv}+(2.703\pm0.309) \nonumber \\ 
\textrm{main$+$high-$z$}\\
\alpha\si{ox}(l\si{uv})=-(0.136\pm0.008) l\si{uv}+(2.616\pm0.249) \nonumber \\ 
\textrm{main$+$high-$z$$+$Sy\,1}
\end{eqnarray}
(the excess precision quoted is useful for plotting purposes). The
residual scatter around the linear relations is 0.14 in $\log$ units
for all samples. The $\alpha$\si{ox}--$l$\si{uv} slopes for all
samples are consistent with those inferred from the
$l$\si{x}--$l$\si{uv} regressions in \S~\ref{luvlx}.

Comparison with previous work is not entirely straightforward, since
the sample selections, X-ray detection fractions, pattern censoring,
and control of other systematics in previous studies differ
substantially from those presented here. \citet{Wilkes94} obtain
$\alpha$\si{ox}--$l$\si{uv} slopes ranging from $-0.1$ to $-0.2$ for
various AGN subsamples, selected from a heterogeneous and incomplete
sample of 343 AGNs, the majority of which were optically selected and
observed with \emph{Einstein}. For a subsample of 272 RQ,
$l\si{uv}>29.0$ AGNs, \citet{Wilkes94} find
$\alpha\si{ox}\propto-(0.15\pm0.03)l\si{uv}$ (see their Figure~14a),
which is consistent with Eqns.(4)--(6) above.
\citet{Green95} use a stacking technique to obtain an $\alpha$\si{ox}--$l$\si{uv} 
relation for 908 Large Bright Quasar Survey AGNs with RASS coverage,
only 10\% of which have X-ray detections.  Binning in luminosity and
redshift, and assuming no redshift dependence, they obtain
$\alpha\si{ox}\propto-(0.08\pm0.02)l\si{uv}$, which is consistent with
our results within 3$\sigma$, but the comparison is inappropriate
since their sample includes both RL and BAL AGNs. The corresponding
slope for the $\alpha$\si{ox}--$l$\si{uv} relation found by VBS03 and
updated by \citet{VignaliHighZ} is $-0.095\pm0.021$ for the SDSS EDR
sample, significant at the \hbox{3--4$\sigma$} level. The higher
significance of the $\alpha$\si{ox}--$l$\si{uv} anti-correlation found
in our new sample is a result of the increased monochromatic
luminosity and redshift coverage, as well as the increased X-ray
detection fraction; the 2$\sigma$ difference in the
$\alpha$\si{ox}--$l$\si{uv} slope is probably caused by the higher
fraction of X-ray upper limits in the VBS03 sample \citep[\about 50\%
in VBS03 and][]{VignaliHighZ}. Aside from the higher statistical 
significance of our current results, we also consider them to be less
prone to systematic errors of the type described in \S~\ref{intro}.

Based on the \hbox{$\alpha\si{ox}-\alpha\si{ox}(l\si{uv})$} residuals,
we can estimate the maximum possible residual dependence of
$\alpha\si{ox}$ on redshift and the corresponding maximum possible
variation of the ratio of UV-to-X-ray flux,
$r=F_{\nu}(\textrm{2500\,\AA})/F_{\nu}(\textrm{2\,keV})$.  Using
 the Kaplan-Meier estimator means of the
\hbox{$\alpha\si{ox}-\alpha\si{ox}(l\si{uv})$} residuals in nine redshift
bins (see inset plot in the top panel of Figure~\ref{luv_z}), we
obtain the weighted linear regression
\hbox{$\left<\alpha\si{ox}-\alpha\si{ox}(l\si{uv})\right>=$}
\hbox{$(0.005\pm0.012)z+(-0.010\pm0.023)$}. The
slope is consistent with zero, which again indicates that there is no
need for an additional redshift dependence. According to the above
linear regression, we expect $\alpha\si{ox}$ to vary by no more than
0.03 between the redshifts of 0 and 5. By definition,
$r=F_{\nu}(\textrm{2500\,\AA})/F_{\nu}(\textrm{2\,keV})=10^{2.606\alpha\si{ox}}$,
and differentiating this with respect to $\alpha$\si{ox}, we have
$\delta
r/r=2.606(\delta\alpha\si{ox})\log_e(10)\approx6(\delta\alpha\si{ox})\approx0.2$,
for $\delta\alpha\si{ox}=0.03$.  This implies that the ratio of
rest-frame UV-to-X-ray flux could only change by $\lesssim$20\% with
cosmic time from $z\approx0-5$. Similar analysis applied to the
\hbox{$\alpha\si{ox}-\alpha\si{ox}(z)$} residuals (see inset plot 
in the right panel of Figure~\ref{luv_z}) confirms that redshift alone
cannot be responsible for the observed variation in $\alpha\si{ox}$. The
\hbox{$\alpha\si{ox}-\alpha\si{ox}(z)$} residuals show a systematic 
variation of \about0.2 between monochromatic luminosities $l\si{uv}=28.5$
and $l\si{uv}=31.8$.

\begin{figure}
\epsscale{1.0}
\plotone{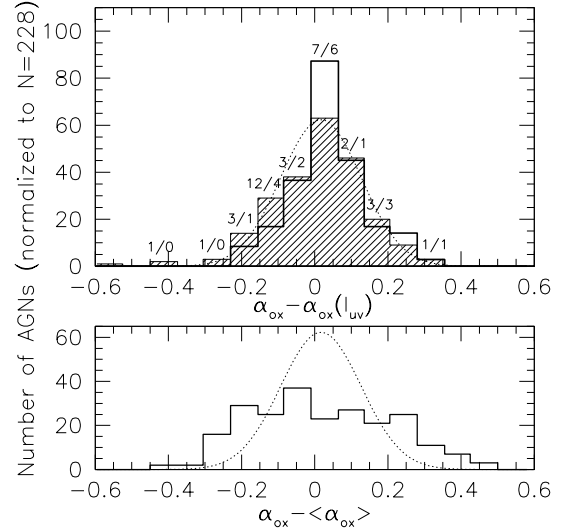}
\caption{Distributions of the $\alpha\si{ox}$ residuals after 
adjusting for the $l\si{uv}$ dependence (top) or the average
$\alpha\si{ox}$ (bottom, assuming no $l\si{uv}$ or $z$
dependence). The hatched histogram in the top panel represents the 228
AGNs from the combined sample; the open histogram represents the 81
AGNs with $1.55<z<4.8$ (normalized to $N=228$ for comparison), for
which all BAL AGNs can be identified using the SDSS spectroscopy. The
dotted Gaussian shown in both panels is centered at
\hbox{$\alpha\si{ox}-\alpha\si{ox}(l\si{uv})=0.017$} and has a standard
deviation of 0.11. The two numbers on the tops of some histogram bins
show the number of AGNs in each bin with X-ray limits in the combined
sample (left number) and the $1.55<z<4.8$ subsample (right number, no
normalization was applied).
\label{a_oxRes}}
\end{figure}

Figure~\ref{a_oxRes} shows the distributions of
$\alpha\si{ox}-\alpha\si{ox}(l\si{uv})$ residuals, adjusted for the
luminosity dependence of $\alpha\si{ox}$ (using Eqn.~6), for both the
combined sample and a $1.55<z<4.8$ subsample (for which all HiBALs can
be identified using SDSS spectroscopy). Both distributions have been
rescaled to $N=228$, the total number of AGNs in the combined
sample. The slight tendency of the combined-sample distribution
towards more negative $\alpha\si{ox}-\alpha\si{ox}(l\si{uv})$ values
is probably a result of the \about9--15 unidentified BALs which
remain in the sample due to lack of \ion{C}{4} spectroscopic coverage
for $z<1.55$ and $z>4.8$. Proper comparison (i.e., one that takes into
account the upper limits) of the two distributions with Gehan and
logrank tests from ASURV shows that they are indistinguishable,
implying that our combined sample does not contain more than a few
percent obscured or X-ray weak AGNs.  The dotted curve in the top
panel of Figure~\ref{a_oxRes} is a Gaussian representation of the
combined-sample residuals with a mean of 0.017 and a standard
deviation of 0.11 (compared to 0.14 obtained from the linear
regression of the combined sample). The Gaussian parameterization
provides a reasonable representation of the residuals in both the
observed (shown in Figure~\ref{a_oxRes}) and the binned differential
Kaplan-Meier distributions. It is unlikely that we can determine
whether a different parametric distribution (e.g., a Lorentzian) will
provide a better fit, since the tails of the distribution are
uncertain due to the small number of objects.  There is no evidence of
significant skewness of the $\alpha\si{ox}$ distribution, after
correction for the luminosity dependence of $\alpha\si{ox}$. If a
significant number of obscured AGNs remained in our sample, we would
see an extended leftward tail of the
$\alpha\si{ox}-\alpha\si{ox}(l\si{uv})$ residuals \citep[if the
absorbed AGNs had X-ray detections, as in Figure~1 of][]{BLW}, or a
significant skewness of the distribution if only upper limits were
available for the BAL AGNs. We suspect that the skewness of the
$\alpha\si{ox}$ distribution seen by \citet[][see their Figure~1, with
the $x$-axis reversed]{AT86} is a result of the presence of obscured
(and possibly a larger fraction of RL) AGNs in their sample. The
bottom panel of Figure~\ref{a_oxRes} presents the
$\alpha\si{ox}-\left<\alpha\si{ox}\right>$ residuals, where
$\left<\alpha\si{ox}\right>=-1.514$ is the Kaplan-Meier average of the
combined sample, assuming no $l$\si{uv} and no redshift
dependence. The broad distribution is a result of ignoring the
$\alpha$\si{ox}--$l$\si{uv} anti-correlation in a sample with a large
range of luminosities.
 
Eqns.~4--6 show that, within the quoted uncertainties, the same slope
and intercept for the $\alpha$\si{ox}--$l$\si{uv} relation are present
for the main, main$+$high-$z$, and combined samples. As detailed in
\S~\ref{luvlx}, these parameter estimates are also unaffected by the
exclusion of the 14 \emph{XMM-Newton}/\emph{Chandra} detected AGNs,
the four \emph{ROSAT} targets, the \hbox{9--15} steepest
$\alpha$\si{ox} AGNs (at the appropriate redshifts) to check for any
effect of the unidentified HiBALs, an additional 10 AGNs with some UV
absorption, or the 17 AGNs considered dust-reddened by the
\citet{Richards03} criterion. The strength of the correlations is
slightly lower (\hbox{7.2--9.4$\sigma$} level for the different
samples) if the 14 \emph{XMM-Newton}/\emph{Chandra} AGNs are excluded,
since this decreases the sample size by \about 7\% and the detection
fraction by \about 1--2\%. If we do not correct for the host-galaxy
contamination in low-luminosity AGNs from the main SDSS sample, Eqn.~4
above would have a somewhat shallower slope of $-0.128\pm0.014$ and an
intercept of $2.377\pm0.417$. The effect is in the expected direction
(taking into account the artificial increase in $l$\si{uv} and
steepening of $\alpha$\si{ox} for the affected AGNs), and its size
($\lesssim1\sigma$) is determined by the fact that only
\about 17\% of the 155 SDSS AGNs have host-galaxy correction
$>$5\%. Even if all Sy~1 AGNs need similar host galaxy corrections,
their effect on the $\alpha$\si{ox}--$l$\si{uv} anti-correlation
parameters will be equally small, as they represent only 16\% of the
full sample (37/228).

\begin{figure}
\epsscale{1.0}
\plotone{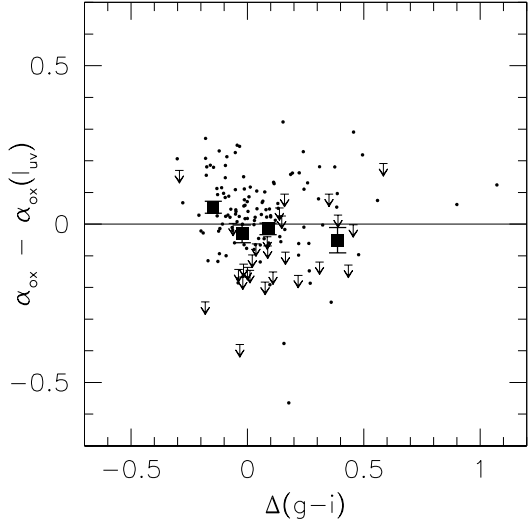}
\caption{Single-parameter fit residuals for an $\alpha$\si{ox} dependence 
on $l\si{uv}$ (from Eqn.~6) vs. $\Delta(g-i)$ for the main SDSS sample
(solid circles). The four $z>3.1$ main-sample AGNs with $\Delta(g-i)$
values affected by the Lyman limit are excluded. Arrows indicate X-ray
upper limits. The Kaplan-Meier estimators of the mean
$\alpha\si{ox}-\alpha\si{ox}(l\si{uv})$ in four $\Delta(g-i)$ bins,
given with large filled squares, show that
$\alpha\si{ox}-\alpha\si{ox}(l\si{uv})$ is not strongly correlated
with $\Delta(g-i)$ over the range where we have coverage.
\label{gi_a_ox_res}}
\end{figure}

Figure~\ref{gi_a_ox_res} presents the
$\alpha\si{ox}-\alpha\si{ox}(l\si{uv})$ residuals of the main SDSS
sample vs. the redshift-corrected \gM$-$\iM\ color, $\Delta(g-i)$.
Although the redder SDSS AGNs with $\Delta(g-i)>0$ appear to be more
likely to have $\alpha\si{ox}$ limits rather than detections
(partially because they have fainter \iM~magnitudes; see
Figure~\ref{sdssBQS}), no trend of the Kaplan-Meier estimators of the
mean $\alpha\si{ox}-\alpha\si{ox}(l\si{uv})$ residuals is apparent
when we bin the data in four $\Delta(g-i)$ bins (selected to have
equal numbers of objects). A Spearman test on the individual data
points returns a correlation coefficient of $-$0.14 with an 8\%
probability of the null hypothesis (no correlation) being correct. We
conclude that any dust-reddening dependence of $\alpha\si{ox}$ (in
addition to the $l\si{uv}$ dependence) must be weak for the main SDSS
sample, at least over the $\Delta(g-i)$ range where we have
significant source statistics.

\subsection{Is the $\alpha$\si{ox}--$l$\si{uv} Relation Non-linear?}

Some studies of optical/UV and X-ray emission from AGNs suggest a
possible non-linear dependence of $\alpha$\si{ox} on $l$\si{uv}
\citep{Wilkes94,Anderson03}. \citet{Wilkes94} observe that the 
$\alpha$\si{ox}--$l$\si{uv} correlation found for the \emph{Einstein}
quasar database, $\alpha\si{ox}\propto-0.11l\si{uv}$, has a flatter
slope, $\alpha\si{ox}\propto-0.08l\si{uv}$, if the sample is
restricted to low-luminosity objects with $l\si{uv}<29.5$. While the
authors cannot rule out a non-linear relation, they suggest that the
difference in slopes is likely caused by the varying host-galaxy
contribution to the $l$\si{uv} measurement at low redshift (which is
accompanied, as expected, by a larger scatter in $\alpha$\si{ox}).
\citet{Anderson03} also report an observed tendency toward a non-linear 
$\alpha$\si{ox}--$l$\si{uv} relation (note that they also use the term
``non-linear'' to refer to the fact that the slope of the linear
$l$\si{x}--$l$\si{uv} relation is less than one). The
\citet{Anderson03} sample contains \about 1158 bright \emph{ROSAT} 
All Sky Survey (RASS) selected AGNs with broad-line SDSS
counterparts. This sample is not optically selected; in fact it
provides X-ray fluxes for only \about 10\% of all SDSS AGNs, the
majority of which are at low redshifts ($z<1$). The goal of the 
\citet{Anderson03} paper was to present the first installment of a 
RASS-SDSS catalog; consequently the presented analysis of the
$\alpha$\si{ox}--$l$\si{uv} relation, as stated by the authors, was
not intended to be conclusive. The effects not taken into account
include sample selection biases, the statistical method which did not
consider third-variable dependencies or the effect of unidentified
BALs, and the effects of the varying dispersions in the optical/UV and
X-ray bands (see \S~\ref{yuan_sec}). To our knowledge, there is
presently no conclusive evidence for a non-linear
$\alpha$\si{ox}--$l$\si{uv} correlation.

From Figure~\ref{a_ox_luv}, it appears that the
$\alpha$\si{ox}--$l$\si{uv} correlation may be non-linear, with a
flatter slope for $l\si{uv}\lesssim30.5$ and a steeper one at higher
monochromatic luminosities. We checked this by performing linear
regressions separately for two subsamples, separated at
$l\si{uv}=30.5$. The results shown below are based on the
main$+$high-$z$ samples, excluding the Sy\,1 sample which is not
optically selected; the combined sample gives qualitatively the same
results. We obtain a slope of $-0.09\pm0.02$ for the $l\si{uv}<30.5$
subsample, and $-0.13\pm0.02$ for the $l\si{uv}>30.5$ subsample. It
appears that the slopes are different at the \about 2$\sigma$ level.
From Figure~\ref{a_ox_luv}, the main SDSS sample has five outlier
points at low monochromatic luminosities (with $l\si{uv}<29$ and
$\alpha\si{ox}<-1.46$), which could have influenced the anti-correlation
found for $l\si{uv}<30.5$ AGNs. If we exclude those points and repeat
the analysis, we obtain slopes of $-0.12\pm0.02$ and $-0.13\pm0.02$
for the $l\si{uv}<30.5$ and $l\si{uv}>30.5$ subsamples, respectively,
implying that the difference in slopes is likely an artifact of the
addition of the five outlier AGNs rather than demonstrating a real
difference. The five outliers are all nearby AGNs, with $z<0.22$, and
most of them are probably X-ray absorbed Seyferts. Exclusion of the
five outlier AGNs has a $1\sigma$ effect on the regression parameters
in the combined sample, steepening the slope from $\alpha\si{ox}
\propto (-0.14\pm0.01) l\si{uv}$ to $\alpha\si{ox} \propto
(-0.15\pm0.01) l\si{uv}$. We conclude that the present sample does not
offer significant evidence for a non-linear
$\alpha$\si{ox}--$l$\si{uv} relation.

\subsection{Validating the Slope of the $l\si{x}$--$l\si{uv}$ Relation}
\label{yuan_sec}

\citet[][C83]{Chanan83}, \citet[][F95]{LaFranca}, and \citet[][YSB98]{Yuan98}
explore the possibility that the intrinsic
\hbox{$l\si{x}$--$l\si{uv}$} relation has a slope of one. They propose
that a larger dispersion in the rest-frame UV (relative to the X-ray)
measurements, combined with the steep bright-end UV luminosity
function, conspire to produce an \hbox{$l\si{x}$--$l\si{uv}$} relation
with a slope smaller than one and an apparent
\hbox{$\alpha$\si{ox}--$l\si{uv}$} correlation.  Both F95 and YSB98
assume Gaussian distributions of uncertainties independent of
luminosity or redshift for $l\si{x}$ and $l\si{uv}$. They take the
observed dispersion around a linear \hbox{$l\si{x}$--$l\si{uv}$}
relation to be \hbox{\about 0.4--0.5} in $\log$ units, corresponding
to a dispersion of \hbox{0.15--0.2} in the $\alpha$\si{ox}--$l\si{uv}$
relation. This is presumably caused by dispersion in the optical/UV
and X-ray measurements due to measurement error, variability, and
intrinsic dispersion (related to differences in accretion modes and
the conditions in the immediate AGN environment as well as the galaxy
host). In order to fit their \emph{Einstein} data with a linear
$l\si{x}$--$l\si{uv}$ relation, F95 require a dispersion in the
rest-frame monochromatic UV luminosity of $\sigma\si{uv}\approx0.34$ in
$\log$ units (corresponding to 0.85\,mag); the known causes of
uncertainty in their sample (i.e., optical/UV photometric measurement
error, assumed constant optical/UV spectral slope, and AGN
variability) account for only 0.5\,mag. Thus, the F95 conclusions
depend on the assumption that the extra scatter observed around the
linear \hbox{$l\si{x}$--$l\si{uv}$} relation is due to extra
dispersion in the optical/UV.

YSB98 and \citet{Yuan99} also assume that the observed dispersion in
the $l\si{x}$--$l\si{uv}$ relation is largely due to Gaussian
uncertainty in the optical/UV. In the notation of \citet{Yuan99},
given \emph{intrinsic} monochromatic luminosities of $\bar{l}\si{x}$ and
$\bar{l}\si{uv}$ modified by (measurement-error, variability, and
intrinsic) scatters of $\delta l\si{x}$ and $\delta l\si{uv}$, the
\emph{observed} monochromatic luminosities are $l\si{x}=\bar{l}\si{x} +
\delta l\si{x}$ and $l\si{uv}=\bar{l}\si{uv} + \delta l\si{uv}$.
The scatters $\delta l\si{x}$ and $\delta l\si{uv}$ are assumed to be
independent of luminosity and redshift and well represented by
Gaussian distributions with zero means and standard deviations of
$\sigma\si{x}$ and $\sigma\si{uv}$.  YSB98 and \citet{Yuan99} caution
that a spurious \hbox{$\alpha$\si{ox}--$l\si{uv}$} relation could
arise for samples with large optical/UV dispersions (with
optical/UV-to-X-ray dispersion ratio,
$R_{\sigma}=\sigma\si{uv}/\sigma\si{x}>1$) and \emph{intrinsic} bright
monochromatic luminosity limits of $\bar{l}^{\textrm{max}}\si{uv}=31.5$.  In
their scenario, the steep bright-end luminosity function produces an
effective bright $l\si{uv}$ cutoff, which together with the large
optical/UV dispersion distorts the \hbox{$l\si{x}$--$l\si{uv}$}
distribution, inducing an apparent correlation with slope smaller than
one (see Figure~5a of YSB98).  Assuming a maximum \emph{observed}
monochromatic luminosity limit of $l\si{uv}^{\textrm{max}}\approx33$
(corresponding to the most powerful AGNs found in many surveys), the
\emph{intrinsic} monochromatic luminosity limit
$\bar{l}^{\textrm{max}}\si{uv}$ is fainter by $\Delta l\si{uv}$, i.e.,
$l\si{uv}^{\textrm{max}}=\bar{l}^{\textrm{max}}\si{uv}+\Delta
l\si{uv}$, with $\Delta l\si{uv}$ given by Eqn. B3 of \citet{Yuan99}:
\begin{equation}
\Delta l\si{uv}=\frac{(\gamma_o-1)(\ln10) R^2_{\sigma} (2.605 \sigma_{\alpha\si{ox}})^2}{1+R^2_{\sigma}}=\sigma^2\si{uv}(\gamma_o-1)\ln10.
\end{equation}
Here $\sigma_{\alpha\si{ox}}$ is the standard deviation of the
observed dispersion around the linear
\hbox{$\alpha$\si{ox}--$l\si{uv}$} relation and $\gamma_o$ is the
slope of the optical luminosity function ($\propto L^{\gamma_o}$,
$\gamma_o=3-4$). From Eqn.~7, a large $\sigma\si{uv}$ combined with a
steep bright-end luminosity-function slope (larger $\gamma_o$) can
cause a large difference between the observed and intrinsic maximum
monochromatic luminosity (large
\hbox{$\Delta l\si{uv}$}) and bias the $l\si{x}$--$l\si{uv}$ slope.
As defined in \citet{Yuan99}, $\sigma\si{uv}$ is related to
$\sigma_{\alpha\si{ox}}$ and $R_{\sigma}$ by:
\begin{equation}
\sigma_{\alpha\si{ox}}=0.3838\,\sqrt{\sigma\si{uv}^2 + \sigma\si{x}^2}=0.3838\,\sigma\si{uv}\sqrt{1+\frac{1}{R^2_{\sigma}}}
\label{sigEqn}
\end{equation}
For a given observed $\sigma_{\alpha\si{ox}}$, larger
optical/UV-to-X-ray dispersion ratios $R_{\sigma}$ are equivalent to a
larger fraction of the observed dispersion being attributed to the
dispersion in the $l\si{uv}$ measurement, $\sigma\si{uv}$, and
potentially larger bias affecting the the \hbox{$l\si{x}$--$l\si{uv}$}
correlation. F95, YSB98, and \citet{Yuan99} take the observed
$\sigma_{\alpha\si{ox}}$, estimate the dispersion in the X-ray
measurements, and assign the remaining observed dispersion to the
rest-frame UV band, assuming no intrinsic X-ray dispersion. Since the
estimated $\sigma\si{x}$ was typically much less than the observed
dispersion around the linear \hbox{$\alpha\si{ox}$--$l\si{uv}$} fit,
$R_{\sigma}>1$ (see Eqn.~\ref{sigEqn}) gives rise to an
$l\si{x}$--$l\si{uv}$ correlation with slope less than one and an
apparent $\alpha\si{ox}$--$l\si{uv}$ correlation. In the following
subsections we consider the sources of dispersion in both the
rest-frame UV and X-ray monochromatic luminosities and confirm that the
$l\si{x}$--$l\si{uv}$ correlation has a slope of
\about 0.65 for all realistic $R_{\sigma}$ values in our sample.

\subsubsection{Dispersion of the $l\si{x}$ and $l\si{uv}$ measurements}
\label{dispersions}
The dispersions of the $l\si{x}$ and $l\si{uv}$ estimates, assuming no
intrinsic dispersion, i.e., $l\si{uv}=A l\si{x}+C$, where $A$ and $C$
are constants independent of monochromatic luminosity or redshift, can
be attributed to measurement errors and AGN variability. AGN
variability is a function of both wavelength and AGN luminosity, and
it affects our results since the optical/UV and X-ray observations are
not simultaneous. For our sample, the \emph{ROSAT} observations were
taken between 1991 and 1993, the \emph{Chandra} and \emph{XMM-Newton}
observations between 2000 and 2002, and the SDSS observations between
2000 and 2003; the timescales of interest are thus of order
\hbox{0--12\,years} (corresponding to rest-frame time lags of
\hbox{0--12\,years}). The optical/UV variability structure function
of AGNs shows signs of flattening for time lags of $>5$\,years, at a
value of \about 0.3\,mag for measurements at 2500\,\AA\ of a typical
SDSS AGN with an absolute $i$-band magnitude $M_i$\about$-25$
\citep{Ivezic04,AGNvar}. A 2500\,\AA\ variability amplitude of 0.3\,mag
corresponds to \about 30\% uncertainty in
$F_{\nu}(\textrm{2500\,\AA})$ and $L\si{uv}$, and \about 4\%
uncertainty in $\alpha$\si{ox}. The measurement uncertainties in the
rest-frame UV are typically \about 10\%, but could be as large as 25\%
for about one-quarter of the main sample, as discussed in
\S~\ref{optical}. If we weight the measurement uncertainties by
the number of AGNs affected, we arrive at an average rest-frame UV
measurement error of \about 14\%. Adding the uncertainties due to
variability and measurement error in quadrature, we expect $\delta
L\si{uv}\sim33\%$ ($\delta l\si{uv}\sim0.17$ in $\log$ units).

The X-ray flux measurements are considerably less certain, with
typical measurement errors of 30\% ($\gtrsim$10--40\% for
$-14<\log(F\si{0.5--2\,keV})<-13$). On short timescales more luminous
AGNs have smaller X-ray variability amplitudes
\citep[e.g.,][]{Green93}, but all AGNs have comparable amplitude
variations on the longer timescales (of order years) of interest to
us. Longer timescale variability studies of Seyfert~1s reveal
variability of $\gtrsim$100\% of the mean count rate in some sources,
with no obvious difference in the variability amplitude between higher
and lower luminosity AGNs \citep[][and references
therein]{Uttley4agn,Uttley04}. Typical long-term root mean square
(rms) variability of Seyfert~1s is 20--40\%
\citep{GrupeVar,Uttley4agn,Markowitz03}. Assuming the long-term variability is
the same in luminous AGNs (\about 30\%), and combining the
uncertainties due to variability and measurement errors, we arrive at
an average uncertainty of $\delta L\si{x}\sim42$\% ($\delta
l\si{x}\sim0.23$ in $\log$ units) for our X-ray measurements.
 
Taking into account only the measurement errors and variability
effects on the $l\si{x}$ and $l\si{uv}$ estimates, we infer X-ray and
optical/UV uncertainties (in log units) of $\sigma\si{x}>0.23$ and
$\sigma\si{uv}<0.17$, respectively. Combining the above estimates, we
arrive at an expected dispersion of \about0.29. The observed
dispersion varies between 0.35 and 0.39 for our combined and main
samples, implying that, unless we are underestimating the uncertainty
due to measurement error and/or variability, there is an extra source
of dispersion roughly equal in magnitude to the one we can account for
that is perhaps intrinsic to the AGN energy generation mechanism. 

\subsubsection{Effect of the $l\si{x}$ and $l\si{uv}$ uncertainties on the measured relations}

\begin{figure}
\epsscale{2.0}
\plottwo{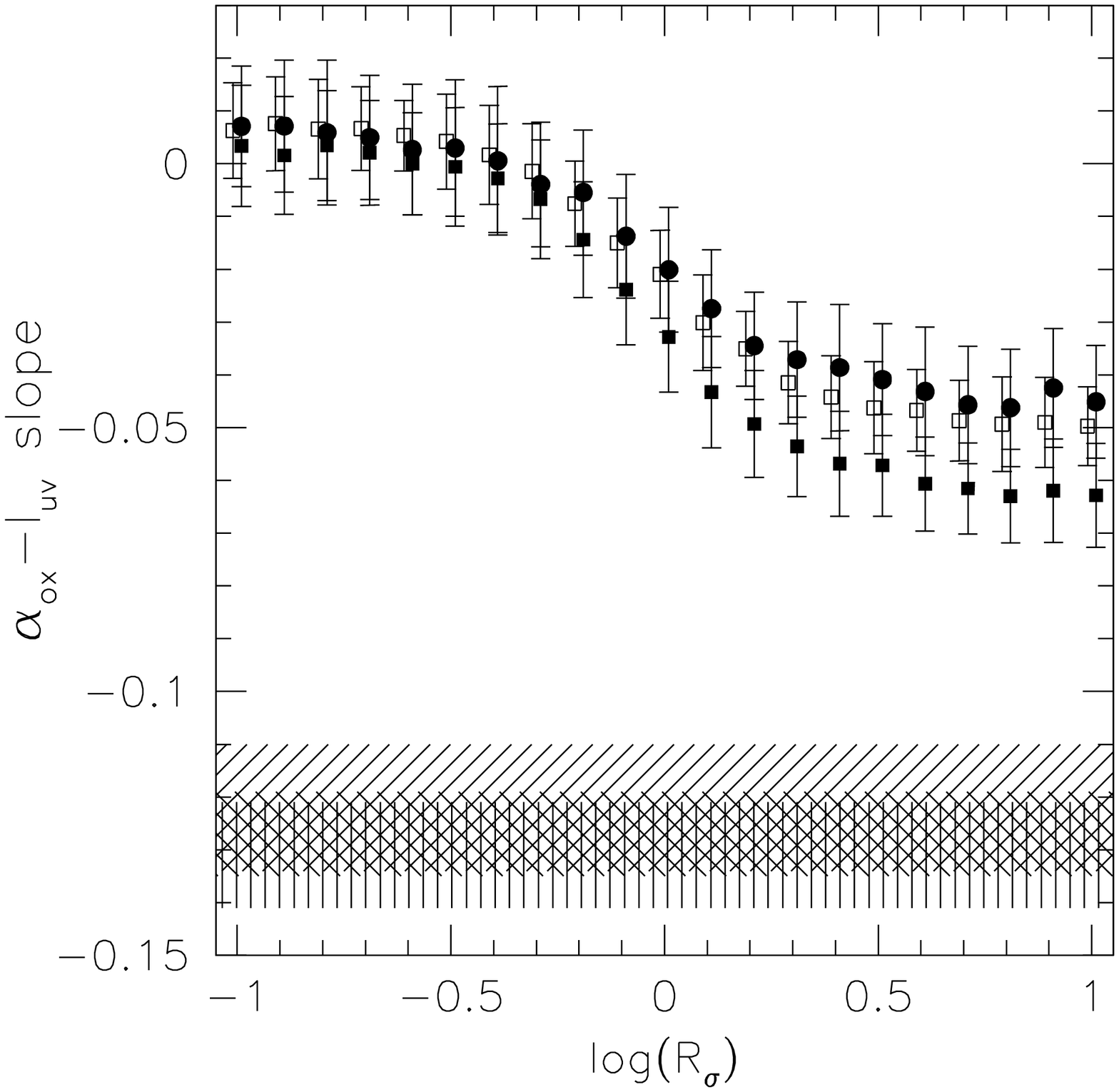}{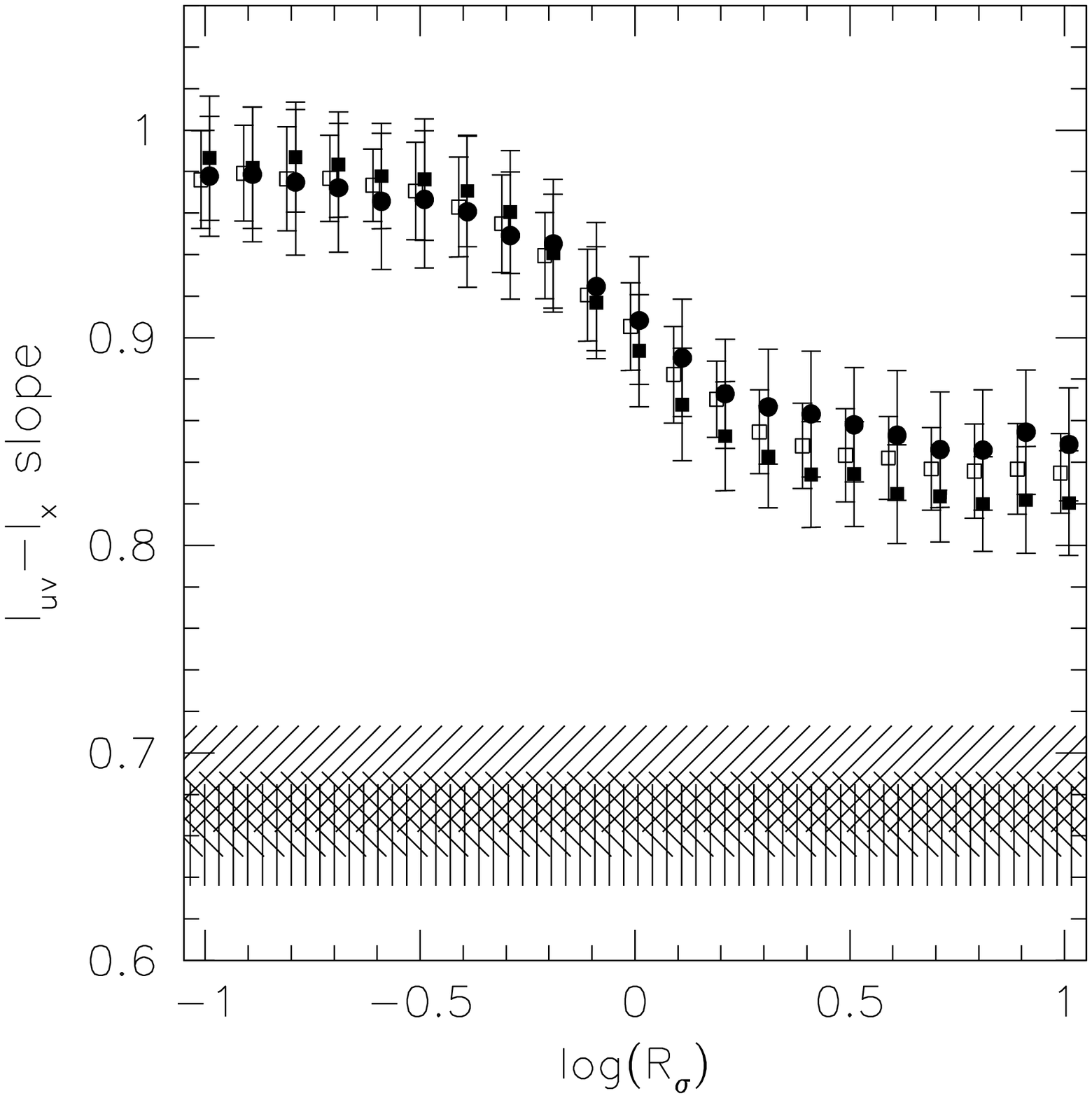}
\caption{Slopes of the apparent $\alpha$\si{ox}-$l\si{uv}$ anti-correlation 
(top) and the $l\si{x}$-$l\si{uv}$ correlation (bottom) as a function
of the optical/UV-to-X-ray dispersion ratio, $R$\si{$\sigma$}, from
simulated samples similar to the main SDSS sample (solid circles), the
main$+$high-$z$ samples (solid squares), and the combined sample (open
squares). Each point represents the average slope obtained from 100
simulated samples, equally spaced in $\log$($R$\si{$\sigma$}); the
squares have been displaced by $\pm$0.01 from the true
$\log$($R$\si{$\sigma$}) values for clarity. The hatched regions
represent the 1$\sigma$ ranges measured in the main SDSS sample
(right-slanted), the main$+$high-$z$ samples (vertical lines), and the
combined sample (left-slanted).
\label{YuanMC}}  
\end{figure}

In the previous section we estimated the dispersions in the $l\si{x}$
and $l\si{uv}$ measurements considering measurement errors and AGN
variability. Here we use Monte Carlo simulations of mock samples to
assess the validity of the sample correlations in the presence of
large dispersion in the rest-frame UV relative to the X-ray band.
From \S~\ref{dispersions}, $\sigma\si{x}\sim0.23$, and the observed
dispersion in $\alpha\si{ox}$ for the main sample is
$\sigma_{\alpha\si{ox}}=0.15$. Even if all the extra dispersion in
$\alpha$\si{ox} comes from the rest-frame UV, $\sigma\si{uv}<0.31$ and
$R_{\sigma}<1.4$ (in log units $\log(R_{\sigma})<0.15$). We simulated
100 samples similar to the main, main$+$high-$z$, and combined samples
(equal numbers of objects with the same rest-frame UV
monochromatic luminosity distribution and equal numbers of X-ray limits) for
each of 21 different $R_{\sigma}$ values, equally spaced in log units
between $\log(R_{\sigma})=-1$ and $\log(R_{\sigma})=1$. For each
$R_{\sigma}$, we computed the average slopes of the
$l\si{x}$--$l\si{uv}$ and $\alpha$\si{ox}--$l\si{uv}$ correlations
from the 100 mock samples of each of our three subsamples (mock-main,
mock-main$+$high-$z$, and mock-combined) and display the results in
Figure~\ref{YuanMC}. None of the ratios of optical/UV-to-X-ray
dispersion considered here can produce an apparent $\alpha\si{ox}$
relation or a $l\si{x}$--$l\si{uv}$ relation with slopes equal to
those observed in the main, the main$+$high-$z$, or the combined
samples with $>99$\% confidence ($>4\sigma$). Our sample estimates
indicate that $\log(R_{\sigma})<0.15$, which only increases the
significance of this comparison. Larger optical/UV-to-X-ray dispersion
ratios than the one considered here are unrealistic, and thus we
conclude that the correlations found in this paper are not apparent
correlations caused by the steep bright end of the optical/UV
luminosity function and a large dispersion in the optical/UV relative
to the X-rays.

\section{Discussion and Conclusions}
\label{conclude}

The SDSS is providing one of the largest optically selected AGN
samples to date with substantially better photometry and higher
completeness than previous well-studied optical color selected samples
like the BQS sample. Various studies have found that
the bright $B$ band selection limit ($B<16.16$) and blue $U-B$ cut
($U-B<-0.44$) of the BQS sample bias the sample towards $z<0.5$ and the
bluest luminous AGNs, systematically excluding redder objects, while
including some AGNs fainter than the quoted magnitude limit 
\citep[e.g.,][]{WP85,W2000,Jester}. SDSS uses 4-dimensional 
redshift-dependent color selection and flux limits the AGN sample in
the \iM-band (with an effective wavelength of 7481\,\AA\, compared to
4400\,\AA\ for the BQS sample's $B$ band), which, together with the
accurate CCD photometry, creates a highly complete, representative
sample of optical AGNs.
 
We have selected a representative sample of 155 radio-quiet SDSS AGNs
from DR2, serendipitously observed in medium-deep \emph{ROSAT}
pointings, creating an unbiased sample with sensitive coverage in the
rest-frame UV, 20\,cm radio, and soft X-ray bands.  Using the
serendipitous \emph{ROSAT} observations of SDSS AGNs supplemented by
36 high-redshift luminous QSOs and 37 Seyfert~1 galaxies, we consider
the relations between rest-frame UV (measured at 2500\,\AA) and X-ray
(at 2\,keV) emission in a combined sample of 228 AGNs with an X-ray
detection fraction of $86$\%.  We have carefully dealt with a variety
of selection and analysis issues, ranging from the appropriateness of
the sample to the suitability of the statistical methods. The removal
of RL and BAL AGNs is essential if we want to study the intrinsic
relations between UV and X-ray energy generation in the typical
luminous AGN, as it restricts the confusing effects of jet emission
and X-ray absorption. To the extent that we can measure them, BAL AGNs
have the same underlying X-ray emission properties as normal RQ AGNs
\citep[e.g.,][]{Gallagher}, but they remain hidden by strong absorption.
Consequently we take special care to remove all known BALs from our
sample and to consider the effects of unidentified BAL AGNs in
specific redshift ranges.

We find that the monochromatic luminosity at 2500\,\AA\ and 2\,keV are
correlated (at the 11.5$\sigma$ level), independent of their strong
correlations with redshift.  This correlation cannot be caused by the
steep fall-off of the bright AGN number-density combined with a large
ratio of optical/UV-to-X-ray dispersion in our sample as suggested by
C83, F95, YSB98, and \citet{Yuan99}. We take special care when
evaluating the statistical significance of partial correlations in
censored datasets.  Using the partial Kendall's $\tau_{12,3}$ and the
EM linear regression method in an optically selected sample with a
wide range of AGN luminosities and redshifts and a large X-ray
detection fraction, we can properly assess the significance and
estimate the parameters of the correlations. In addition, we use Monte
Carlo realizations of mock relations in simulated samples to establish
the applicability of the above methods.  We confirm that the slope of
the $l\si{x}$--$l\si{uv}$ correlation is less than one (\about 0.65),
implying a dependence of the optical/UV-to-X-ray index,
$\alpha$\si{ox}, on monochromatic luminosity and/or redshift. We find that
$\alpha$\si{ox} is primarily dependent on rest-frame monochromatic UV
luminosity (at the 7.4--10.6$\sigma$ level), while any redshift
dependence is insignificant ($\lesssim1.1\sigma$). 

The $\alpha$\si{ox}--$l$\si{uv} anti-correlation implies that AGNs
redistribute their energy in the UV and X-ray bands depending on
overall luminosity, with more luminous AGNs emitting fewer X-rays per
unit UV luminosity than less luminous AGNs. Currently, no
self-consistent theoretical study is able to explain from first
principles why $\alpha$\si{ox} should be in the observed range, much
less predict its variation with $l$\si{uv}. Theoretical studies of
\citet{SS73} disks give predictions of the rest-frame UV emission but
cannot predict the X-ray emission, which is believed to originate in a
hot coronal gas of unknown geometry and disk-covering fraction. Recent
advances in magnetohydrodynamic simulations of accretion disks
\citep[e.g.,][and references therein]{BH98} offer the promise of a
self-consistent disk$+$corona model of AGN emission. In such a model,
the dissipation of magnetic fields, arising from the
magneto-rotational instability deep in the accretion disk, could heat
the coronal gas to X-ray emitting temperatures \citep[J.~H. Krolik
2004, private communication; see also][]{Krolik99}.  Our empirical
relation between rest-frame UV and soft-X-ray emission in AGNs and the
$\alpha\si{ox}$--$l$\si{uv} anti-correlation provide the best
constraints yet that future self-consistent disk$+$corona models must
explain.
 
The observed lack of redshift dependence of $\alpha$\si{ox} at fixed
luminosity provides evidence for the remarkable constancy of the
accretion process in the immediate vicinity of the black hole, despite
the dramatic changes of AGN hosts and the strong evolution of AGN
number densities over the history of the Universe. The sample used
here provides no evidence for non-linearities in the
$\alpha$\si{ox}--$l$\si{uv} relation. The dispersions observed around
the $l\si{x}$--$l\si{uv}$ and $\alpha$\si{ox}--$l\si{uv}$ relations
cannot be accounted for by measurement errors and AGN variability
alone, suggesting that black-hole mass, accretion rate, and/or spin
(and the corresponding differences in accretion modes, energy
generation mechanisms, and feedback) could be contributing to the
observed dispersion.

Our results are qualitatively consistent with previous studies
\citep[e.g.,][]{AT86,VBS03}, but the new results are quantitatively
better since they are based on a large, highly complete sample with
medium-deep soft X-ray coverage and carefully controlled systematic
biases.  Although larger samples of optically selected AGNs with X-ray
coverage can be constructed
\citep[e.g.,][]{Wilkes94,Green95,Anderson03}, the existing survival 
analysis tools cannot guarantee an accurate recovery of the intrinsic
rest-frame UV to X-ray relations based on pattern censored data with
shallow X-ray coverage and low X-ray detection fraction. Stacking
analysis can be used on optical AGNs with shallow X-ray coverage
\citep[e.g.,][]{Green95}, but this method provides only mean values, 
without constraining the spread in each bin. In addition, stacking
analyses done to date have not always allowed for binning in Galactic
Hydrogen column densities, redshifts, radio-loudness, and strong
UV-absorption.  The $l\si{x}$--$l\si{uv}$ relation presented here can
be used to predict more accurately the intrinsic X-ray fluxes of AGNs
with known optical/UV luminosity and serves to define the ``normal''
range of soft X-ray emission for a typical AGN (i.e., RQ, non-BAL
AGNs, unaffected by absorption). Based on this definition of normal
X-ray emission, it is easier to determine if a ``special'' class of
AGNs differs in its X-ray properties from normal AGNs. X-ray ``weak''
AGNs are an example of such a special AGN class. \citet{Risaliti} used
the BQS sample to define normal AGNs, and suggested that some AGNs in
the Hamburg Quasar Survey \citep[HQS,][]{hqs} are X-ray weaker in
comparison. However, \citet{BrandtSDSS} caution that since the HQS
AGNs are among the most luminous objects in the rest-frame UV, the
observed steep $\alpha$\si{ox} values are expected based on the
$\alpha$\si{ox}--$l\si{uv}$ anti-correlation for about half of the
objects (see their Figure~3). Our more accurate prediction of the
optical/UV-to-X-ray emission of normal AGN will also allow researchers
to constrain the X-ray emission associated with jets in RL AGNs
\citep[assuming that AGN jets do not contribute to the emission at
2500\,\AA, but see][]{BH95a,BH95b,3C279} and to study the X-ray
properties of other special AGNs; e.g, red AGNs, AGNs without emission
lines, or AGNs with unusual emission lines \citep[e.g.,][]{exotic}.
The $\alpha$\si{ox}--$l$\si{uv} relation of normal AGNs presented in
this paper can also lead to more accurate estimates of the bolometric
luminosities of AGNs, resulting in tighter constraints on the
importance of AGN-phase mass accretion for the growth of supermassive
black holes as described in, e.g., \citet{Marconi}. Assuming the
\citet{elvis94} spectral energy distribution (SED) and
$-1.7<\alpha\si{ox}<-1.26$ (where the majority of our optically selected
RQ non-absorbed AGNs lie; see Figure~\ref{a_ox_luv}) together with the
$\alpha$\si{ox}--$l$\si{uv} relation from Eqn.~6, we estimate that the
ratio of the \hbox{0.5--2.0\,keV} luminosity to the bolometric
luminosity varies by a factor of 6--9 over the luminosity range
$l\si{uv}=28.5-31.8$ (depending on the inclusion or exclusion of the
infrared bump in the computation of the bolometric luminosity). If
neglected, the variation of the bolometric correction with AGN
luminosity could lead to substantial systematic errors in bolometric
luminosity estimates.

Future SDSS data releases will allow the enlargement of the
optical/UV/soft-X-ray sample of AGNs, as well as provide large new
samples of optically selected AGNs serendipitously observed with
\emph{XMM-Newton} and \emph{Chandra}, as the sky-coverage of X-ray 
satellites increases with time. Larger samples will include more
homogeneous low-luminosity AGN data, providing more sensitive
constraints on the non-linearity of the
\hbox{$\alpha$\si{ox}--$l\si{uv}$} relation.  In addition,
longer-wavelength optical/UV monochromatic flux estimates would complement
the rest-frame UV measurements at 2500\,\AA\ used here, to minimize
any effects of dust absorption in the UV on the
\hbox{$l\si{x}$--$l\si{uv}$} relation \citep[e.g.,][but see also
Hopkins et al.\,2004]{Gaskell}.  The extension to samples observed in
harder X-ray bands is also necessary to constrain the possible effects
of soft-X-ray absorption better. This can be achieved by considering
an $\alpha$\si{ox} index computed using rest-frame $\approx$5\,keV
instead of 2\,keV X-ray monochromatic fluxes.

\citet{Hasinger} reports that X-ray selected AGN samples have 
$l\si{x}$--$l\si{uv}$ correlations consistent with a slope of one and
no $\alpha$\si{ox} dependence on either luminosity or
redshift. Current X-ray selected samples with optical identifications
are large and cover wide ranges of optical/UV and X-ray luminosity,
but they seldom constrain the optical/UV absorption, radio loudness,
or host-galaxy contribution of the sources. In addition, some X-ray
selected samples are biased toward particular optical AGN types
\citep[e.g., narrow-line Seyfert~1s in bright soft X-ray samples;][]{Grupe} 
and could contain a larger fraction of absorbed AGNs. More studies are
necessary to reconcile the results obtained for optically
color-selected and X-ray selected samples, taking into account the
sample selection effects in flux limited samples introduced by the
optical/UV and X-ray AGN population number density and luminosity
evolution with cosmic time.

Firmly establishing the correlation between rest-frame UV and X-ray
emission in AGNs is the first step toward understanding their
generation mechanisms and interrelations. A reasonable next step is to
try to relate the correlations found here to reasonable estimates of
black-hole masses and accretion rates. The difficulty in this endeavor
lies in the fact that direct black-hole mass measurements and
bolometric luminosity estimates are not available for large AGN
samples like those considered here. Indirect black-hole mass
measurements can be obtained from a combination of monochromatic
luminosity and broad emission-line width measurements as shown for BQS
sample AGNs by \citet{Kaspi} and SDSS AGNs by \citet{MD04}. Such
estimates, however, will depend on the extrapolation of \citet{Kaspi}
relation from lower ($\lambda L_{\lambda}(5100\,\textrm{\AA})\lesssim
2 \times 10^{45}$, corresponding to $l\si{uv}\lesssim30.3$) to higher
($\lambda L_{\lambda}(5100\,\textrm{\AA})\gtrsim 2 \times 10^{45}$)
luminosity AGNs, the use of different emission lines at different
redshifts (e.g., H$\beta$ and \ion{Mg}{2} for the SDSS sample
presented here), as well as a non-trivial correction for the effects
of the host-galaxy, giving rise to possible systematic errors.  We are
currently investigating the feasibility of this endeavor for the SDSS
sample presented here.

\vskip 0.5in \leftline{Acknowledgments} We thank M. Akritas for
fruitful discussions on the suitability of the statistical methods
used in the paper and P.~Hall and J.~R.~Trump for help with the BAL
classification of SDSS AGNs. The authors acknowledge the help of
W.~Yuan in implementing his work on the slope validation of the
$l\si{x}$--$l\si{uv}$ correlation and the help of David Schlegel with
SDSS spectroscopy. We are grateful to Gordon Richards, Sebastian
Jester and Sarah Gallagher, as well as an anonymous referee, for
helping us to improve this manuscript. IVS, WNB, and DPS acknowledge
the support of NASA LTSA grant NAG5-13035 and the CXC grant
GO2-3134X. CV acknowledges the support of MIUR COFIN grant 03-02-23.

Funding for the Sloan Digital Sky Survey (SDSS) has been provided by
the Alfred P. Sloan Foundation, the Participating Institutions, the
National Aeronautics and Space Administration, the National Science
Foundation, the U.S. Department of Energy, the Japanese
Monbukagakusho, and the Max Planck Society. The SDSS Web site is
http://www.sdss.org/.  The SDSS is managed by the Astrophysical
Research Consortium (ARC) for the Participating Institutions. The
Participating Institutions are The University of Chicago, Fermilab,
the Institute for Advanced Study, the Japan Participation Group, The
Johns Hopkins University, Los Alamos National Laboratory, the
Max-Planck-Institute for Astronomy (MPIA), the Max-Planck-Institute
for Astrophysics (MPA), New Mexico State University, University of
Pittsburgh, Princeton University, the United States Naval Observatory,
and the University of Washington.

This research has made use of the NASA/IPAC Extragalactic Database
(NED) which is operated by the Jet Propulsion Laboratory, California
Institute of Technology, under contract with the National Aeronautics
and Space Administration.



\oddsidemargin=-1cm
\tabletypesize{\scriptsize}

\begin{deluxetable}{lccrrrcccccccccc} 
\tablewidth{0pt} 
\tablenum{1}
\tabletypesize{\tiny}
\tablecaption{SDSS-\emph{ROSAT} AGN data} \tablehead{\colhead{SDSS ID} &\colhead{$z$} &\colhead{Obs. ID} &\colhead{$T$\si{exp}} &\colhead{$\delta$} &\colhead{Cnt} &\colhead{$f$\si{2500}} &\colhead{$f$\si{x}} &\colhead{$f$\si{2\textrm{keV}}} &\colhead{$\alpha\si{ox}$} &\colhead{$l$\si{x}} &\colhead{$l$\si{uv}} &\colhead{\iM} &\colhead{$\Delta(g-i)$} &\colhead{Flag}\\ \colhead{(1)} &\colhead{(2)} &\colhead{(3)} &\colhead{(4)} &\colhead{(5)} &\colhead{(6)} &\colhead{(7)} &\colhead{(8)} &\colhead{(9)} &\colhead{(10)} &\colhead{(11)} &\colhead{(12)} &\colhead{(13)} &\colhead{(14)} &\colhead{(15)}}
\startdata
SDSSJ012602.21$-$001924.1 & 1.7659 & rp800645n00 & 12.0 & 11 & 19.0 & $-$3.71 & $-$13.70 & $-$8.08 & $-$1.68 & 26.80 & 31.17 & 17.94 & $+$0.144 & 1 \\
SDSSJ021000.72$-$100358.0 & 1.9604 & rp800114n00 & 13.8 & 9 & 72.0 & $-$4.16 & $-$13.20 & $-$7.56 & $-$1.30 & 27.41 & 30.81 & 19.23 & $-$0.180 & 1 \\
SDSSJ022225.49$-$090258.5 & 0.2242 & rp800016n00 & 11.6 & 14 & 5.6 & $-$4.28 & $-$13.91 & $-$8.65 & $-$1.68 & 24.41 & 28.78 & 18.49 & $-$0.033 & 0 \\
SDSSJ022226.11$-$085701.3 & 0.1667 & rp800016n00 & 11.5 & 15 & 445.4 & $-$4.17 & $-$12.32 & $-$7.08 & $-$1.12 & 25.72 & 28.63 & 17.85 & $+$0.193 & 1 \\
SDSSJ022356.30$-$085707.8 & 1.5762 & rp800016n00 & 12.5 & 10 & 13.0 & $-$4.13 & $-$13.92 & $-$8.34 & $-$1.62 & 26.45 & 30.67 & 18.90 & $+$0.220 & 1 \\
SDSSJ022435.93$-$090001.3 & 1.6118 & rp800016n00 & 9.2 & 19 & 15.1 & $-$4.12 & $-$13.62 & $-$8.03 & $-$1.50 & 26.78 & 30.69 & 19.11 & $-$0.015 & 1 \\
SDSSJ023305.95$+$003856.4 & 0.2441 & rp800482n00 & 26.1 & 11 & 107.0 & $-$5.13 & $-$13.31 & $-$8.04 & $-$1.12 & 25.10 & 28.01 & 18.52 & $+$0.559 & 1 \\
SDSSJ023306.26$+$004614.5 & 2.2906 & rp800482n00 & 26.5 & 9 & 36.2 & $-$4.65 & $-$13.82 & $-$8.13 & $-$1.34 & 26.96 & 30.44 & 20.33 & $-$0.159 & 1 \\
SDSSJ023325.32$+$002914.9 & 2.0171 & rp800482n00 & 25.1 & 16 & 81.1 & $-$3.74 & $-$13.43 & $-$7.77 & $-$1.55 & 27.21 & 31.24 & 18.25 & $+$0.007 & 1 \\
SDSSJ023333.24$+$010333.1 & 2.0587 & rp800482n00 & 19.6 & 19 & 5.5 & $-$3.72 & $-$14.02 & $-$8.37 & $-$1.78 & 26.64 & 31.28 & 18.30 & $-$0.038 & 0 \\
\enddata

\tablecomments{The complete version of this table is in the electronic 
edition of the Journal. The printed edition contains only a sample of
ten objects. Note 1: These AGNs were the targets for their respective
\emph{ROSAT} PSPC pointings. Note 2: These AGNs were not detected in the
selected \emph{ROSAT} pointings; X-ray fluxes are from
\emph{XMM-Newton}, \emph{Chandra}, or shorter \emph{ROSAT} exposures
as specified. The units of luminosity are erg\,s$^{-1}$, of
monochromatic luminosity -- erg\,s$^{-1}$\,Hz$^{-1}$. Columns: (1)
SDSS ID; (2) redshift; (3) X-ray observation ID; (4) the effective
X-ray exposure time, $T$\si{exp}, in 10$^3$sec; (5) $\delta$, the
X-ray source off-axis angle in arcmin; (6) total source counts,
corrected for background and aperture size; the precision quoted is
higher than the accuracy; (7) $f$\si{2500}, the logarithm of the
2500\,\AA\ monochromatic flux, not band-pass corrected; (8)
$f$\si{x}, the logarithm of the 0.5-2\,keV flux, not band-pass
corrected; (9) $f$\si{2\textrm{keV}} the logarithm of the 2\,keV
monochromatic flux, not band-pass corrected; (10) $\alpha\si{ox}$, the
optical/UV-to-X-ray index; (11) $l$\si{x}, the logarithm of the 2\,keV
monochromatic luminosity, band-pass corrected; (12) $l$\si{uv}, the
logarithm of the 2500\,\AA\ monochromatic luminosity, band-pass
corrected; (13) \iM, the point source SDSS apparent magnitude,
corrected for Galactic extinction; (14) the relative PSF color,
$\Delta(g-i)$, corrected for Galactic extinction; (15) this flag is
set to 1 if the AGN is X-ray detected.}
\label{tab1}

\end{deluxetable}

\begin{deluxetable}{crrr} 
\tablewidth{0pt} 
\tablenum{2}
\tablecaption{X-ray Detection Fractions} 
\tablehead{\colhead{Sample} &\colhead{X-ray} &\colhead{Total} &\colhead{\% X-ray} \\ \colhead{} &\colhead{Detected} &\colhead{AGN} &\colhead{Detected}}
\startdata
Main            & 126 & 155 &  81\% \\
High-$z$        &  32 &  36 &  89\% \\
Sy\,1           &  37 &  37 & 100\% \\
Combined        & 195 & 228 &  86\% \\
\enddata
\label{tab2}
\end{deluxetable}

\begin{deluxetable}{llcccrc} 
\tablewidth{0pt} 
\tablenum{3}
\tablecaption{X-ray UV correlations} 
\tablehead{\colhead{Relation} &\colhead{Sample} &\colhead{$N$\si{AGN}} &\colhead{\% X-ray} &\colhead{Signi-} &\colhead{$\tau\si{12,3}$} &\colhead{Regression}\\ \colhead{} &\colhead{} &\colhead{} &\colhead{Detected} &\colhead{ficance} &\colhead{} &\colhead{Residuals}}
\startdata
$l\si{2keV}$ vs. $l\si{2500\,\AA}$      & M,H,S & 228 & 86\% & 11.5$\sigma$ & 0.38      & 0.36 \\
$l\si{2keV}$ vs. $l\si{2500\,\AA}$      & M,H   & 191 & 83\% &  8.7$\sigma$ & 0.30      & 0.37 \\
$l\si{2keV}$ vs. $l\si{2500\,\AA}$      & M     & 155 & 81\% &  7.4$\sigma$ & 0.28      & 0.39 \\ 
$\alpha\si{ox}$ vs. $l\si{2500\,\AA}$   & M,H,S & 228 & 86\% & 10.6$\sigma$ & $-$0.33   & 0.14 \\
$\alpha\si{ox}$ vs. $l\si{2500\,\AA}$   & M,H   & 191 & 83\% &  9.2$\sigma$ & $-$0.32   & 0.14 \\
$\alpha\si{ox}$ vs. $l\si{2500\,\AA}$   & M     & 155 & 81\% &  7.4$\sigma$ & $-$0.30   & 0.15 \\
$\alpha\si{ox}$ vs. $z$                 & M,H,S & 228 & 86\% &  1.1$\sigma$ & $+$0.03   & ...  \\
$\alpha\si{ox}$ vs. $z$                 & M,H   & 191 & 83\% &  1.1$\sigma$ & $-$0.03   & ...  \\
$\alpha\si{ox}$ vs. $z$                 & M     & 155 & 81\% &  1.0$\sigma$ & $-$0.02   & ...  \\ 
\enddata
\tablecomments{Sample M refers to the main SDSS sample, sample H to 
the high-$z$ sample, and sample S to the Sy\,1 sample. All cases test
partial correlations, taking into account the effect of a third
variable which is either redshift (in the first six cases) or
$l\si{2500\,\AA}$ (in the last three).}
\label{tab3}
\end{deluxetable}

\end{document}